\documentclass[%
 reprint, nofootinbib, amsmath,amssymb, aps,floatfix,
]{revtex4-2}
\usepackage{gensymb}
\usepackage{textcomp}
\usepackage{lipsum}
\usepackage{graphicx}
\usepackage{dcolumn}
\usepackage{bm}
\usepackage{siunitx}
\DeclareSIUnit\gauss{G}
\DeclareSIUnit\erg{erg}

\usepackage{cancel}
\usepackage{tabularx}
\usepackage{amssymb}
\usepackage{amsmath}
\usepackage{relsize}
\usepackage{listings}
\usepackage{commath}
\usepackage{enumitem}
\usepackage{xfrac}
\usepackage{float}
\usepackage{rotating}
\usepackage{booktabs}
\usepackage{makecell}
\usepackage{mathtools}
\usepackage{caption}
\usepackage{subcaption}
\usepackage{multirow}
\usepackage[table,xcdraw]{xcolor}
\usepackage[version=4]{mhchem}
\usepackage[colorlinks,bookmarks=false,citecolor=red,linkcolor=blue,urlcolor=blue]{hyperref}

\begin{document}

\preprint{APS/123-QED}
\title{Selection pressure/Noise driven cooperative behaviour in the thermodynamic limit of repeated games}
\author{Rajdeep Tah}
\email{rajdeep.phys@gmail.com}
\author{Colin Benjamin}
\email{colin.nano@gmail.com}
\affiliation{School of Physical Sciences, National Institute of Science Education and Research, HBNI, Jatni-752050, India\\
Homi Bhabha National Institute, Training School Complex, Anushakti Nagar, Mumbai-400094, India}

\begin{abstract}
    Consider the scenario where an infinite number of players (i.e., the \textit{thermodynamic} limit) find themselves in a Prisoner's dilemma type situation, in a \textit{repeated} setting. Is it reasonable to anticipate that, in these circumstances, cooperation will emerge? This paper addresses this question by examining the emergence of cooperative behaviour, in the presence of \textit{noise} (or, under \textit{selection pressure}), in repeated Prisoner's Dilemma games, involving strategies such as \textit{Tit-for-Tat}, \textit{Always Defect}, \textit{GRIM}, \textit{Win-Stay, Lose-Shift}, and others. To analyze these games, we employ a numerical Agent-Based Model (ABM) and compare it with the analytical Nash Equilibrium Mapping (NEM) technique, both based on the \textit{1D}-Ising chain. We use \textit{game magnetization} as an indicator of cooperative behaviour. A significant finding is that for some repeated games, a discontinuity in the game magnetization indicates a \textit{first}-order \textit{selection pressure/noise}-driven phase transition, particularly when strategies do not punish opponents for defecting even once. We also observe that in these particular cases, the phase transition critically depends on the number of \textit{rounds} the game is played in the thermodynamic limit. For all five games, we find that both ABM and NEM, in conjunction with game magnetization, provide crucial inputs on how cooperative behaviour can emerge in an infinite-player repeated Prisoner's dilemma game.
\end{abstract}

\maketitle

\section{\label{introsec}Introduction}
In the evolutionary context, when we think about examples of social dilemmas, the first thing that comes to our mind is the Prisoner's dilemma \cite{ref1, ref2}. It is one of the most popular game theoretic models that can be used to study a vast array of topics, involving both \textit{one-shot} and \textit{repeated} game settings (see, Refs.~\cite{ref3, ref4, ref5, ref7, ref8}). To recall, \textit{one-shot} games involve a single round of decision-making by the players with no subsequent interactions, while \textit{repeated} games entail multiple iterations of the same game played over time, allowing for strategic adaptation, by the players, based on past actions and future expectations \cite{ref2a}. In prisoner's dilemma, as the word ``\textit{dilemma}" in the name suggests, the Nash equilibrium, i.e., a set of \textit{actions} (or, \textit{strategies}) that lead to an outcome deviating from which one gets worse payoffs, is the \textit{Defect} strategy. This is surprising since there exists a \textit{Pareto optimal} outcome, which for prisoner's dilemma, has better payoffs for both the players and is associated with the \textit{Cooperate} strategy. The \textit{Pareto optimal} strategy and the \textit{Nash equilibrium} strategy are not the same. However, \textit{defection} is not always a viable strategy for players in the Prisoner's dilemma game, especially in a repeated game setting \cite{ref2a}, since the threat of future punishment (due to \textit{defection}) by other players incentivizes \textit{cooperation}. This is because cooperative strategies yield higher long-term benefits through repeated interactions and reputation effects, as revealed by studying Prisoner's dilemma games \cite{ref1, ref2, ref2a}. Till now, most research papers (see, Refs.~\cite{ref3, ref9, ref10, ref11, ref12, ref19}) dealing with thermodynamic limit have been largely restricted to the \textit{one-shot} prisoner's dilemma. However, much less focus has been given to the \textit{repeated} Prisoner's dilemma game in the \textit{thermodynamic} (or, infinite player) limit.

\textit{Repeated} prisoner's dilemma\cite{ref1, ref2}, as the name suggests, is a slightly modified version of the \textit{one-shot} prisoner's dilemma in which the players play the prisoner's dilemma game in a ``\textit{repeated}" environment rather than in a ``\textit{one-shot}" setting. In some previous works (see, Refs.~\cite{ref12, ref12a}), ``\textit{one-shot}" games were considered, i.e., the players made decisions without any knowledge of each other's past actions/strategies. In \textit{one-shot} games, there was no possibility for players to observe each other's behaviour over time or to adjust their strategies based on past outcomes, i.e., each player typically makes a single decision, and the game concludes after that decision is made. In \textit{repeated} prisoner's dilemma, the ``\textit{action}" of a player signifies the player's decision at a particular round in the game, and a sequence of ``\textit{actions}" relates to the ``\textit{strategy}" that the player employs when engaging in the game. In a \textit{repeated} game, if the players respond to the \textit{action} taken by their opponent in the preceding round, then \textit{reactive} tactics such as \textit{Always cooperate} (All-C), i.e., the player will always cooperate; \textit{Always defect} (All-D), i.e., the player will always defect; \textit{Tit-for-Tat} (TFT), i.e., the TFT player will copy its opponent's strategy; \textit{GRIM}, i.e., the players playing \textit{GRIM} cooperates until the opponent defects, after which the player defects indefinitely; \textit{GRIM*}, a slight modification of \textit{GRIM} wherein \textit{GRIM*} players will definitely \textit{defect} in the last round due to a lack of incentive to cooperate in the last round; and \textit{Win-stay-Loose-shift} (WSLS), i.e., player strategies that lead to positive outcomes are more likely to be repeated in the future -- \textit{win-stay}, while strategies that lead to negative outcomes are more likely to be changed in the future -- \textit{lose-shift}, can be played\cite{ref6a, ref6c}.

Previously, it was shown, in Refs.~\cite{ref6a, ref6b, ref6c}, that the study of \textit{repeated} games is pivotal for several reasons. Firstly, it provides insights into the \textit{emergence} and \textit{sustainability} of cooperative behaviours among rational individuals/players, a phenomenon often elusive in \textit{one-shot} interactions\cite{ref2a}. By allowing for the accumulation of \textit{reputation} and the possibility of \textit{direct reciprocity} (see, Ref.~\cite{ref1, ref2} for more details), \textit{repeated} games offer a more realistic lens through which to understand and predict cooperative strategies in various social and economic settings. Moreover, \textit{repeated} games in the thermodynamic limit address certain limitations inherent in \textit{two}-player and \textit{finite}-player games. In traditional game theory models, the focus is primarily on isolated interactions between a fixed number of players. However, this fails to capture the complexities arising from repeated interactions and the potential for strategic adaptation over time. On the other hand, \textit{repeated} games, played in the thermodynamic limit, offer a richer framework that better encapsulates the dynamics of real-world interactions, including considerations of \textit{reputation}, \textit{trust}, and long-term \textit{incentives}\cite{ref6a, ref6b}.

In this paper, we will study and analyze, both numerically and analytically, how cooperative behaviour arises among an \textit{infinite} number of players playing a \textit{repeated} Prisoner's dilemma game. To do so, we will use a game order parameter, namely, game magnetization $\mu$ (which is analogous to its thermodynamic counterpart, i.e., magnetization). We will adopt the numerical Agent-based modelling (ABM) technique to study cooperative behaviour among players in the \textit{repeated} Prisoner's dilemma game, and we will compare our results with the analytical Nash equilibrium mapping (NEM) method. There exist other analytical methods, like Darwinian selection (DS) and Aggregate selection (AS), to analyze $\mu$, in addition to the NEM method. However, in Ref.~\cite{ref12}, we have shown via a detailed calculation the incorrectness of these analytical methods. Hence, in this work, we will only compare the NEM with the numerical ABM, since both DS and AS are incorrect. Both ABM and NEM are based on the \textit{1D}-Periodic Ising chain (or, IC) with nearest neighbour interactions (see, Refs.~\cite{ref6, ref7, ref10, ref12}). Before moving further, we try to understand what game magnetization actually means. In a symmetric \textit{2-player, 2-strategy} social dilemma game, 2 players (say, $\mathfrak{P}_1$ and $\mathfrak{P}_2$) have 2 different strategies $\$_1$ and $\$_2$ available at their disposal. Thus they have a choice between the two accessible strategies $(\$_1~\text{or}~\$_2)$ which, for each of them, could have varying or identical outcomes (aka, \textit{payoffs}). The \textit{four} strategy sets for $\mathfrak{P}_1$ and $\mathfrak{P}_2$, i.e., $(\$|_{\mathfrak{P}_1},~\$|_{\mathfrak{P}_2}) \in \{(\$_1, \$_1),~ (\$_2, \$_1), ~(\$_1, \$_2),~ (\$_2, \$_2)\}$, are each linked to the payoffs $(\mathrm{m, n, p, q})$ via the \textit{symmetric} payoff matrix ($\Lambda$):
\begin{equation}
    \Lambda = \left[\begin{array}{c|c c}
    	 & \$_1 & \$_2\\ 
    	\hline 
    	\$_1 & \mathrm{m,m} & \mathrm{n,p}\\
        \$_2 & \mathrm{p,n} & \mathrm{q,q}
    \end{array}\right]. 
    \label{eq2.0a}
\end{equation}
\textit{Game magnetization} ($\mu$), analogous to thermodynamic magnetization, is calculated by subtracting the fraction of players choosing $\$_2$ strategy from the fraction of players choosing $\$_1$ strategy. Similar to \textit{one-shot} games \cite{ref12, ref12a}, in repeated games, we introduce the concept of \textit{noise} (or, \textit{selection pressure}), defined as the \textit{uncertainty} associated with a player's strategy selection, denoted by $\beta$.

A brief introduction to \textit{one-shot} and \textit{repeated} prisoner's dilemma, along with a note on NEM and ABM, is presented in Sec.~\ref{theory} and Sec.~\ref{therm} respectively. In Sec.~\ref{theory}, we will discuss five different cases of the \textit{2}-player \textit{repeated} Prisoner's Dilemma, in which both players play either of the strategies: \textit{GRIM}, \textit{TFT}, \textit{All-D}, \textit{GRIM*}, or \textit{WSLS}. In Sec.~\ref{therm}, we will also understand how a \textit{repeated} prisoner's dilemma game can be mapped to the $1D$-Ising chain (that forms the basis for both NEM and ABM) in the \textit{thermodynamic} limit. In \textit{repeated} prisoner's dilemma, we find that for \textit{three} cases: \textit{GRIM} vs. \textit{All-D}, \textit{TFT} vs. \textit{All-D}, and \textit{GRIM} vs. \textit{GRIM*}, ABM and NEM results for game magnetization and its variation with a changing game parameter exactly match each other. However, for the other two cases: \textit{WSLS} vs. \textit{TFT} and \textit{WSLS} vs. \textit{GRIM}, the ABM and NEM results for game magnetization agreed fairly well, although not exactly identical like those derived for the previous three cases. For all cases, except the \textit{GRIM*} vs. \textit{GRIM} and \textit{WSLS} vs. \textit{GRIM} cases, we observe an interesting phenomenon of a \textit{first}-order phase transition, i.e., a change in players' strategies, at a critical payoff value, and the phase transition rate had an intrinsic dependency on both noise $\beta$ as well as the number of rounds the game is played in the thermodynamic limit. For the \textit{GRIM*} vs. \textit{GRIM} case, we observed that a majority of players play the \textit{GRIM*} strategy over the \textit{GRIM} strategy, since \textit{GRIM*} results in a better payoff than \textit{GRIM} in the last round of the game (we discuss this in Sec.~\ref{gg}). However, in the case of \textit{WSLS} vs. \textit{GRIM}, for finite \textit{noise} $\beta$, we observed that all players shift to \textit{GRIM} irrespective of the payoff (\textit{reward} in this case; details in Sec.~\ref{theory}), indicating that \textit{GRIM} becomes the Nash equilibrium. This particular case, where players have to play either \textit{WSLS} or \textit{GRIM}, leads to the same result in both \textit{2}-player and \textit{infinite}-player limits (see, Sec.~\ref{wsgr}). 

The paper is structured as follows: In Sec.~\ref{theory}, we give a brief introduction to \textit{one-shot} and \textit{repeated} prisoner's dilemma, along with discussing the concept of \textit{discount factor} that is associated with \textit{repeated} games. In Sec.~\ref{theory}, we will also give a brief introduction to five different cases of the \textit{2}-player \textit{repeated} Prisoner's Dilemma, in which both players play either of the strategies: \textit{GRIM}, \textit{TFT}, \textit{All-D}, \textit{GRIM*}, or \textit{WSLS}. Further, in Sec.~\ref{therm}, we will discuss the analytical NEM and the numerical ABM techniques that we utilize to study the games in the thermodynamic limit, followed by a detailed analysis of the results obtained for the different cases of the repeated prisoner's dilemma game. Finally, we conclude our paper by summarizing all the important observations from our work in Sec.~\ref{conc-rep}. 


\section{\label{theory}Theory}
In this section, we will first discuss the \textit{one-shot} prisoner's dilemma game, followed by a description of the \textit{discount factor} (which is exclusively defined for \textit{repeated} games) and its role in determining the number of \textit{rounds} for which the game will be played in the \textit{thermodynamic} (or, \textit{infinite} player) limit, followed by a brief introduction on calculating the Nash equilibrium associated with a repeated game in the \textit{2}-player limit. While calculating the Nash equilibrium, we consider \textit{five} deterministic strategies, namely \textit{Always defect} (All-D), \textit{Tit-for-Tat} (TFT), \textit{GRIM}, \textit{GRIM*} (a modified version of \textit{GRIM}), \textit{Win-Stay Lose-Shift} (WSLS) and how they fare against each other.

\subsection{\label{Prisoner's dilemma}Prisoner's Dilemma in the \textit{one}-shot settings}
In the \textit{one-shot} Prisoner's dilemma\cite{ref1, ref2}, two \textit{independent} players (say, $\mathfrak{P}_1$ and $\mathfrak{P}_2$), accused of committing a crime, are being interrogated by the law agencies, and they have either option to \textit{Cooperate} ($\mathfrak{C}$) with each other or \textit{Defect} ($\mathfrak{D}$). If both players opt for $\mathfrak{C}$-strategy, then they are rewarded with a payoff $\mathbb{R}$, whereas, if both choose $\mathfrak{D}$-strategy, then they get the punishment payoff $\mathbb{P}$. However, if both $\mathfrak{P}_1$ and $\mathfrak{P}_2$ choose opposite strategies, then the one choosing $\mathfrak{C}$-strategy gets the \textit{sucker's} payoff $\mathbb{S}$, and the one choosing $\mathfrak{D}$-strategy gets the temptation payoff $\mathbb{T}$, respectively. The payoffs obey the criteria: $\mathbb{T}>\mathbb{R}>\mathbb{P}>\mathbb{S}$. Hence, the Prisoner's dilemma payoff matrix ($\Tilde{\Xi}$) is,
\begin{equation}
    \Tilde{\Xi} = \left[\begin{array}{c|c c}
    	 & \mathfrak{C} & \mathfrak{D}\\ 
    	\hline 
    	\mathfrak{C} & \mathbb{R, R} & \mathbb{S, T}\\
        \mathfrak{D} & \mathbb{T, S} & \mathbb{P, P}
    \end{array}\right]. 
    \label{eq2a.1}
\end{equation}
From $\Tilde{\Xi}$ in Eq.~(\ref{eq2a.1}), one would think that the two rational players, who are always looking for payoff maximization, would choose the Pareto optimal $\mathfrak{C}$-strategy since it is a \textit{win-win} situation for both. However, owing to independence in strategy selection, both $\mathfrak{P}_1$ and $\mathfrak{P}_2$ choose the $\mathfrak{D}$-strategy (thus the name ``\textit{dilemma}") to ensure that none of them receives the minimum payoff (the sucker's payoff $\mathbb{S}$) due to a unilateral change in the opponent's strategy. Hence, the Nash equilibrium in the \textit{one-shot} Prisoner's dilemma is defect $\mathfrak{D}$. However, we are interested to know what happens when the Prisoner's dilemma game is played in a \textit{repeated} game setting. In this scenario, a sequence of cooperation ($\mathfrak{C}$) and defection ($\mathfrak{D}$) \textit{actions} would result in various \textit{strategies} that the players can adopt.

\subsection{\label{df}The discount factor}
In \textit{repeated} games, the \textit{discount factor} $(w)$ signifies the weight assigned to \textit{future} payoffs compared to \textit{immediate} payoffs, reflecting a player's patience\cite{ref2a, ref6b}. Ranging from $0$ to $1$, i.e., $w \in [0, 1]$, $w = 1$ implies equal valuation of \textit{future} and \textit{immediate} payoffs, while $w < 1$ indicates discounting future payoffs. The \textit{discount factor} $w$ can never be greater than $1$ because it ensures that future payoffs are valued less than or equal to immediate payoffs, reflecting the preference for immediate rewards \cite{ref2, ref2a}. The choice of $w$ influences strategy selection, with higher values promoting cooperation as players prioritize long-term gains, while lower values encourage short-term thinking, potentially leading to selfish or competitive behaviours favouring immediate payoffs over future gains\cite{ref18}.

There exists an intricate relation between the discount factor $w$ and the number of rounds for which the game is being played\cite{ref2b}. A higher number of rounds allows players to consider long-term consequences, making cooperative strategies more viable, especially with higher discount factors\cite{ref2a, ref2b}. In shorter games, immediate gains are prioritized, increasing the risk of defection and undermining cooperation, particularly with lower discount factors\cite{ref2a, ref2b}. 

In \textit{repeated} games, the {discount factor} $w$ is multiplied with successive payoffs to account for the diminishing present value of future payoffs. This multiplication reflects the fact that players typically value immediate payoffs more than future payoffs due to factors such as uncertainty, risk, and time preferences. By discounting future payoffs, players effectively weigh them less in their decision-making process, leading to strategic choices that balance short-term gains with long-term considerations\cite{ref2a, ref2b}.



\subsection{\label{2rept}Nash equilibrium for 2-player repeated games}
To have an extensive understanding of how to calculate the Nash equilibrium associated with 2-player \textit{repeated} games, where both players can opt for one of the many different \textit{deterministic} strategies available, we consider \textit{five} unique examples in which players play either \textit{Always defect} (All-D) or \textit{Tit-for-Tat} (TFT) or \textit{GRIM} or \textit{GRIM*} (a modified version of \textit{GRIM}) or \textit{Win-Stay Lose-Shift} (WSLS). A brief description of how \textit{2} players playing \textit{two} of these strategies is given in the following subsections.

\subsubsection{TFT vs. All-D}
\begin{figure}[H]
    \centering
    \includegraphics[width=1\columnwidth]{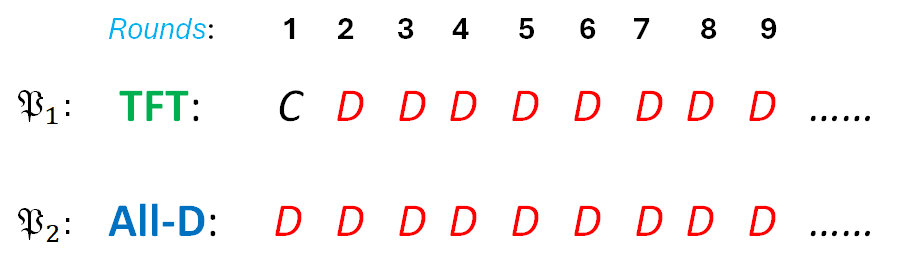}
    \caption{\textit{TFT} vs. \textit{All-D} in the \textit{2}-player limit.}
    \label{fig:2rep}
\end{figure}

We consider a \textit{2}-player \textit{repeated} prisoner's dilemma game, where one of the players (row player; $\mathfrak{P}_1$) plays the Tit-for-Tat (\textit{TFT}) strategy, whereas its opponent (column player; $\mathfrak{P}_2$) plays the Always defect \textit{(All-D)} strategy (see, Fig.~\ref{fig:2rep}). Since \textit{TFT} is a deterministic strategy \cite{ref2a, ref6a}, we can define a fixed \textit{discount factor} $w$ for this game; i.e., the players can determine in advance the number of rounds for which the game will be played (see, Sec.~\ref{df}). We note in Fig.~\ref{fig:2rep} that the player playing \textit{TFT} (or, $\mathfrak{P}_1$ in this case) starts with cooperation $C$ (from the definition of \textit{TFT}; see Sec.~\ref{introsec}) whereas the player playing \textit{All-D} (or, $\mathfrak{P}_2$ in this case) starts with defection $D$.  
Since, $\mathfrak{P}_1$ starts with $C$ while $\mathfrak{P}_2$ starts with $D$ (both play \textit{round} \textbf{1}; see Fig.~\ref{fig:2rep}), $\mathfrak{P}_1$ receives the \textit{sucker's payoff} $\mathbb{S}$ while $\mathfrak{P}_2$ receives the \textit{temptation} $\mathbb{T}$ in the \textit{first} round of the game (see, Eq.~(\ref{eq2a.1}) for \textit{one-shot} prisoner's dilemma). However, for subsequent rounds (i.e., \textit{round} \textbf{2}, \textit{round} \textbf{3}, \textit{round} \textbf{4}, \ldots), both $\mathfrak{P}_1$ and $\mathfrak{P}_2$ play $D$ since $\mathfrak{P}_1$ (who is playing \textit{TFT}) permanently switches to \textit{defection} (a consequence of \textit{direct reciprocity}\cite{ref2a} that exists between the two players) since its opponent always chooses to \textit{defect}. We recall that a \textit{TFT} player mimics its opponent's previous move. This indicates that after the \textit{first} round, both $\mathfrak{P}_1$ and $\mathfrak{P}_2$ receive the punishment payoff $\mathbb{P}$. Hence, we can write the cumulative payoff (denoted by $\Lambda$), coupled with the \textit{discount factor} $w$ defined for subsequent rounds of the game (to account for the decreasing present value of future payoffs over time \cite{ref2b}), for $\mathfrak{P}_1$ and $\mathfrak{P}_2$ as \cite{ref2a},
\begin{gather}
    \Lambda_{\mathfrak{P}_1} = \mathbb{S} + w\mathbb{P} + w^2\mathbb{P} + w^3 \mathbb{P} + \ldots,
    \nonumber\\
    \text{or,}~~\Lambda_{\mathfrak{P}_1} \approx \mathbb{S} + \dfrac{w\mathbb{P}}{1-w},~\text{for large $w$},\label{eq8rep}\\
    \Lambda_{\mathfrak{P}_2} = \mathbb{T} + w\mathbb{P} + w^2\mathbb{P} + w^3 \mathbb{P} + \ldots, \nonumber\\
    \text{or,}~~\Lambda_{\mathfrak{P}_2} \approx \mathbb{T} + \dfrac{w\mathbb{P}}{1-w},~\text{for large $w$}. \label{eq9rep}
\end{gather}
Similarly, if $\mathfrak{P}_1$ played \textit{All-D} while $\mathfrak{P}_2$ played \textit{TFT}, then the payoffs for $\mathfrak{P}_1$ and $\mathfrak{P}_2$ given in Eqs.~(\ref{eq8rep}, \ref{eq9rep}) would just get swapped. 

Now, instead of one playing \textit{TFT} and the other one playing \textit{All-D}, if both the players play \textit{TFT}, it is apparent that both will receive the cumulative payoff (coupled with the \textit{discount factor} $w$; defined for subsequent rounds of the game),
\begin{gather}
    \Lambda_{\mathfrak{P}_1} = \Lambda_{\mathfrak{P}_2} = \mathbb{R} + w\mathbb{R} + w^2 \mathbb{R} + \ldots,\nonumber\\
    \text{or,}~~\Lambda_{\mathfrak{P}_1} = \Lambda_{\mathfrak{P}_2} \approx \dfrac{\mathbb{R}}{1-w},~\text{for large $w$},
    \label{eq10rep}
\end{gather}
since both players will try not to \textit{defect} as the first one to defect will receive the \textit{temptation} payoff $\mathbb{T}$ while its opponent will receive the \textit{sucker's payoff} $\mathbb{S}$ (which is clearly not a viable payoff for \textit{rational} players; both of them look for payoff maximization), and since both are playing \textit{TFT}, if any one of the players \textit{defect} at any round, then both players will receive a periodic combination of $\mathbb{T}$ and $\mathbb{S}$ for the subsequent rounds of the game. One might have noticed that if the game is played for, say, $\Bar{m}$ number of rounds, then each of the players (who are, in this case, playing \textit{TFT}) receives the cumulative payoff of $\Bar{m}\mathbb{R}$ (total number of rounds $\times$ payoff received at each round; in this case, the payoff received at each round is $\mathbb{R}$). Hence, after comparing the cumulative payoff $\Bar{m}\mathbb{R}$ with the cumulative payoff given for each player in Eq.~(\ref{eq10rep}), we find that the number of rounds $\Bar{m}$ is related to the discount factor $w$ via the relation: $\Bar{m} = \frac{1}{1-w}$. This relation can also be derived from Eq.~(\ref{eq8rep}) by considering the fact that for $(\Bar{m}-1)$ number of rounds, both players receive the \textit{punishment} payoff $\mathbb{P}$, i.e., $(\Bar{m}-1)\mathbb{P} = \frac{w\mathbb{P}}{1-w}$, or, $\Bar{m} = \frac{1}{1-w}$.

Similarly, if both the players play \textit{All-D}, then both will receive the cumulative payoff, coupled with the \textit{discount factor} $w$, as,
\begin{gather}
    \Lambda_{\mathfrak{P}_1} = \Lambda_{\mathfrak{P}_2} = \mathbb{P} + w\mathbb{P} + w^2 \mathbb{P} + \ldots,\nonumber\\
    \text{or,}~~\Lambda_{\mathfrak{P}_1} = \Lambda_{\mathfrak{P}_2} \approx \dfrac{\mathbb{P}}{1-w} = \Bar{m}\mathbb{P},~\text{for large $w$}.
    \label{eq11rep}
\end{gather}
Hence, from Eqs.~(\ref{eq8rep}-\ref{eq11rep}), we can derive the payoff matrix $\Lambda$ corresponding to the \textit{two} players who choose either of the two strategies: \textit{TFT} or \textit{All-D}, (\textit{row} player: $\mathfrak{P}_1$, \textit{column} player: $\mathfrak{P}_2$),
\begin{equation}
\small{
    \Lambda = \left[\begin{array}{c|c c} 
    	 & {\textit{TFT}} & {\textit{All-D}}\\ 
    	\hline \\
        {\textit{TFT}} & \dfrac{\mathbb{R}}{1-w}, \dfrac{\mathbb{R}}{1-w} & \mathbb{S} + \dfrac{w\mathbb{P}}{1-w}, \mathbb{T} + \dfrac{w\mathbb{P}}{1-w}\\
        \\
        {\textit{All-D}} & \mathbb{T} + \dfrac{w\mathbb{P}}{1-w}, \mathbb{S} + \dfrac{w\mathbb{P}}{1-w} & \dfrac{\mathbb{P}}{1-w}, \dfrac{\mathbb{P}}{1-w}
    \end{array}\right].}
    \label{eq12rep}
\end{equation}
Since, $\Lambda$ is a \textit{symmetric} matrix, we can equivalently write the \textit{reduced} payoff matrix (we also denote this by $\Lambda$ for brevity), for a single player (the \textit{row} player: $\mathfrak{P}_1$),
\begin{equation}
    \Lambda = \left[\begin{array}{c|c c} 
    	 & {\textit{TFT}} & {\textit{All-D}}\\ 
    	\hline \\
        {\textit{TFT}} & \dfrac{\mathbb{R}}{1-w} & \mathbb{S} + \dfrac{w\mathbb{P}}{1-w}\\
        \\
        {\textit{All-D}} & \mathbb{T} + \dfrac{w\mathbb{P}}{1-w} & \dfrac{\mathbb{P}}{1-w}
    \end{array}\right].
    \label{eq13rep}
\end{equation}
From Eq.~(\ref{eq13rep}), we find that \textit{TFT} becomes the \textit{strict} Nash equilibrium iff ${\Lambda}_{[TFT, TFT]}$ ($\equiv \frac{\mathbb{R}}{1-w}$, in Eq.~\ref{eq13rep}) $>~{\Lambda}_{[All-D, TFT]}$ ($\equiv \mathbb{T} + \frac{w\mathbb{P}}{1-w}$, in Eq.~\ref{eq13rep}), i.e., if $\mathfrak{P}_1$ unilaterally changes from \textit{All-D} to \textit{TFT}, then he receives a better payoff, while the payoff received by $\mathfrak{P}_2$ remains the same. This leads us to the condition: $ \frac{\mathbb{R}}{1-w} > \mathbb{T} + \frac{w\mathbb{P}}{1-w}$ or, $w > \frac{\mathbb{T-R}}{\mathbb{T-P}}$. This shows that for discount factors $w > \frac{\mathbb{T-R}}{\mathbb{T-P}}$, \textit{TFT} becomes the \textit{strict} Nash equilibrium of the game. Otherwise, if $w < \frac{\mathbb{T-R}}{\mathbb{T-P}}$, \textit{All-D} becomes the \textit{strict} Nash equilibrium.

\subsubsection{GRIM vs. All-D}
\begin{figure}[H]
    \centering
    \includegraphics[width=1\columnwidth]{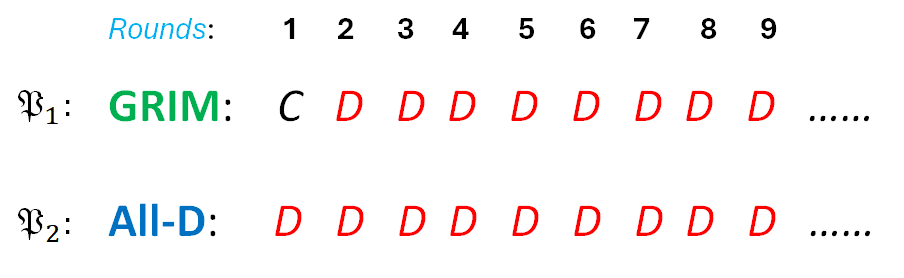}
    \caption{\textit{GRIM} vs. \textit{All-D} in the \textit{2}-player limit.}
    \label{fig:1rep}
\end{figure}

Let us assume that in a \textit{2}-player \textit{repeated} prisoner's dilemma game, one of the players (row player; $\mathfrak{P}_1$) plays the \textit{GRIM} strategy, whereas its opponent (column player; $\mathfrak{P}_2$) plays the Always defect \textit{(All-D)} strategy (see, Fig.~\ref{fig:1rep}). Since both \textit{GRIM} and \textit{All-D} are deterministic strategies, we can define a fixed \textit{discount factor} $w$ for this game; i.e., the players can determine in advance the number of rounds for which the game will be played (see, Sec.~\ref{df}). We also note in Fig.~\ref{fig:1rep} that the player playing \textit{GRIM} (or, $\mathfrak{P}_1$ in this case) starts with cooperation $C$ (from the definition of \textit{GRIM}) whereas the player playing \textit{All-D} (or, $\mathfrak{P}_2$ in this case) starts with defection $D$.  
Since, $\mathfrak{P}_1$ starts with $C$ while $\mathfrak{P}_2$ starts with $D$ (both play \textit{round} \textbf{1}; see Fig.~\ref{fig:1rep}), $\mathfrak{P}_1$ receives the \textit{sucker's payoff} $\mathbb{S}$ while $\mathfrak{P}_2$ receives the \textit{temptation} $\mathbb{T}$ in the \textit{first} round of the game (see, Eq.~(\ref{eq2a.1}) for \textit{one-shot} prisoner's dilemma). However, for subsequent rounds (i.e., \textit{round} \textbf{2}, \textit{round} \textbf{3}, \ldots), both $\mathfrak{P}_1$ and $\mathfrak{P}_2$ play $D$ since $\mathfrak{P}_1$ (who is playing \textit{GRIM}) permanently switches to \textit{defection} (a consequence of \textit{direct reciprocity}\cite{ref2a} that exists between the two players). This indicates that after the \textit{first} round, both $\mathfrak{P}_1$ and $\mathfrak{P}_2$ receive the punishment payoff $\mathbb{P}$. Hence, we can write the cumulative payoff (denoted by $\Lambda$), coupled with the \textit{discount factor} $w$ defined for subsequent rounds of the game (to account for the decreasing present value of future payoffs over time), for $\mathfrak{P}_1$ and $\mathfrak{P}_2$ as\cite{ref2a},
\begin{gather}
    \Lambda_{\mathfrak{P}_1} = \mathbb{S} + w\mathbb{P} + w^2\mathbb{P} + w^3 \mathbb{P} + \ldots,\nonumber\\
    \text{or,}~~\Lambda_{\mathfrak{P}_1} \approx \mathbb{S} + \dfrac{w\mathbb{P}}{1-w},~\text{for large $w$ (via \textit{GP sum})}\label{eq2rep}\\
    \Lambda_{\mathfrak{P}_2} = \mathbb{T} + w\mathbb{P} + w^2\mathbb{P} + w^3 \mathbb{P} + \ldots, \nonumber\\
    \text{or,}~~\Lambda_{\mathfrak{P}_2} \approx \mathbb{T} + \dfrac{w\mathbb{P}}{1-w},~\text{for large $w$ (via \textit{GP sum})}. \label{eq3rep}
\end{gather}
Similarly, if $\mathfrak{P}_1$ played \textit{All-D} while $\mathfrak{P}_2$ played \textit{GRIM}, then the payoffs for $\mathfrak{P}_1$ and $\mathfrak{P}_2$ given in Eqs.~(\ref{eq2rep}, \ref{eq3rep}) would just get swapped. The result obtained for the payoffs, in Eqs.~(\ref{eq2rep}, \ref{eq3rep}), are exactly the same to that of the payoffs obtained for both the players in the \textit{TFT} vs. \textit{All-D} case (see, Eqs.~(\ref{eq8rep}, \ref{eq9rep})). 

Now, instead of one playing \textit{GRIM} and the other one playing \textit{All-D}, if both the players play \textit{GRIM}, it is apparent that both will receive the cumulative payoff (coupled with the \textit{discount factor} $w$; defined for subsequent rounds of the game) as,
\begin{gather}
    \Lambda_{\mathfrak{P}_1} = \Lambda_{\mathfrak{P}_2} = \mathbb{R} + w\mathbb{R} + w^2 \mathbb{R} + \ldots,\nonumber\\
    \text{or,}~~\Lambda_{\mathfrak{P}_1} = \Lambda_{\mathfrak{P}_2}\approx \dfrac{\mathbb{R}}{1-w},~\text{for large $w$ (via \textit{GP sum})},
    \label{eq4rep}
\end{gather}
since both players will try not to \textit{defect} (even once!) as the first one to defect will receive the \textit{temptation} payoff $\mathbb{T}$ while its opponent will receive the \textit{sucker's payoff} $\mathbb{S}$ (which is clearly not a viable payoff for \textit{rational} players; both of them look for payoff maximization). Here too, we notice that if the game is played for, say, $\Bar{m}$ number of rounds, then each of the players (who are, in this case, playing \textit{GRIM}) receives the cumulative payoff of $\Bar{m}\mathbb{R}$ (total number of rounds $\times$ payoff received at each round; in this case, the payoff received at each round is $\mathbb{R}$). Hence, after comparing the cumulative payoff $\Bar{m}\mathbb{R}$ with the cumulative payoff given for each player in Eq.~(\ref{eq4rep}), we find that the number of rounds $\Bar{m}$ is related to the discount factor $w$ via the relation: $\Bar{m} = \frac{1}{1-w}$. This relation can also be derived from Eq.~(\ref{eq2rep}) by considering the fact that for $(m-1)$ number of rounds, both players receive the \textit{punishment} payoff $\mathbb{P}$, i.e., $(m-1)\mathbb{P} = \frac{w\mathbb{P}}{1-w}$, or, $\Bar{m} = \frac{1}{1-w}$.

Similarly, if both the players play \textit{All-D}, then both will receive the cumulative payoff, coupled with the \textit{discount factor} $w$, as,
\begin{gather}
    \Lambda_{\mathfrak{P}_1} = \Lambda_{\mathfrak{P}_2} = \mathbb{P} + w\mathbb{P} + w^2 \mathbb{P} + \ldots,\nonumber\\
    \text{or,}~~\Lambda_{\mathfrak{P}_1} = \Lambda_{\mathfrak{P}_2}\approx \dfrac{\mathbb{P}}{1-w} = m\mathbb{P},~\text{for large $w$}.
    \label{eq5rep}
\end{gather}
Hence, from Eqs.~(\ref{eq2rep}-\ref{eq5rep}), we can derive the payoff matrix $\Lambda$ corresponding to the \textit{two} players who choose either of the two strategies: \textit{GRIM} or \textit{All-D}, (\textit{row} player: $\mathfrak{P}_1$, \textit{column} player: $\mathfrak{P}_2$),
\begin{equation}
\small{
    \Lambda = \left[\begin{array}{c|c c} 
    	 & {\textit{GRIM}} & {\textit{All-D}}\\ 
    	\hline \\
        {\textit{GRIM}} & \dfrac{\mathbb{R}}{1-w}, \dfrac{\mathbb{R}}{1-w} & \dfrac{w\mathbb{P}}{1-w} + \mathbb{S}, \dfrac{w\mathbb{P}}{1-w} + \mathbb{T} \\
        \\
        {\textit{All-D}} & \mathbb{T} + \dfrac{w\mathbb{P}}{1-w}, \mathbb{S} + \dfrac{w\mathbb{P}}{1-w} & \dfrac{\mathbb{P}}{1-w}, \dfrac{\mathbb{P}}{1-w}
    \end{array}\right].}
    \label{eq6rep}
\end{equation}
Since, $\Lambda$ is a \textit{symmetric} matrix, we can equivalently write the \textit{reduced} payoff matrix (we also denote this by $\Lambda$ for brevity), for a single player (the \textit{row} player: $\mathfrak{P}_1$),
\begin{equation}
    \Lambda = \left[\begin{array}{c|c c} 
    	 & {\textit{GRIM}} & {\textit{All-D}}\\ 
    	\hline \\
        {\textit{GRIM}} & \dfrac{\mathbb{R}}{1-w} & \mathbb{S} + \dfrac{w\mathbb{P}}{1-w}\\
        \\
        {\textit{All-D}} & \mathbb{T} + \dfrac{w\mathbb{P}}{1-w} & \dfrac{\mathbb{P}}{1-w}
    \end{array}\right].
    \label{eq7rep}
\end{equation}
From Eq.~(\ref{eq7rep}), we find that \textit{GRIM} becomes the \textit{strict} Nash equilibrium (i.e., a strategy profile where no player can unilaterally deviate for a better payoff) iff ${\Lambda}_{[GRIM, GRIM]}$ ($\equiv \frac{\mathbb{R}}{1-w}$, in Eq.~\ref{eq7rep}) $>~{\Lambda}_{[All-D, GRIM]}$ ($\equiv \mathbb{T} + \frac{w\mathbb{P}}{1-w}$, in Eq.~\ref{eq7rep}), i.e., if $\mathfrak{P}_1$ unilaterally changes from \textit{All-D} to \textit{GRIM}, then he receives a better payoff, while the payoff received by $\mathfrak{P}_2$ remains the same. This leads us to the condition: $ \frac{\mathbb{R}}{1-w} > \mathbb{T} + \frac{w\mathbb{P}}{1-w}$ or, $w > \frac{\mathbb{T-R}}{\mathbb{T-P}}$. This shows that for discount factors $w > \frac{\mathbb{T-R}}{\mathbb{T-P}}$, \textit{GRIM} becomes the \textit{strict} Nash equilibrium of the game. Otherwise, if $w < \frac{\mathbb{T-R}}{\mathbb{T-P}}$, \textit{All-D} becomes the \textit{strict} Nash equilibrium.

From the payoff matrix $\Lambda$ given in Eqs.~(\ref{eq13rep}) and (\ref{eq7rep}), it is evident that both \textit{TFT} vs. \textit{All-D} and \textit{GRIM} vs. \textit{All-D} have identical payoff matrices and therefore in subsequent analysis, we discuss \textit{TFT} vs. \textit{All-D} and \textit{GRIM} vs. \textit{All-D} as a single case.

\subsubsection{\label{gg}GRIM vs. GRIM*}
\begin{figure}[!ht]
    \centering
    \includegraphics[width=1\linewidth]{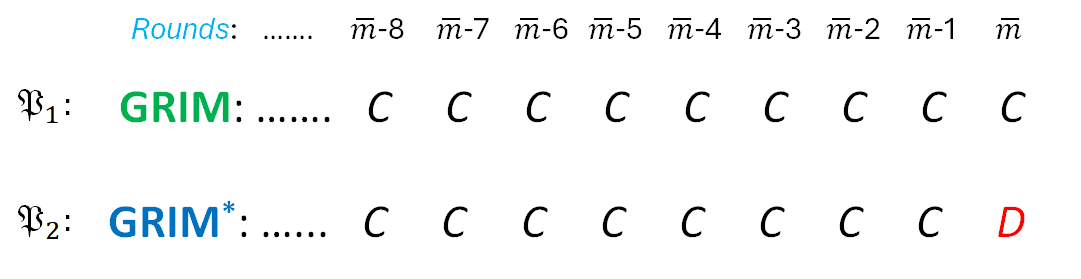}
    \caption{\textit{GRIM} vs. \textit{GRIM*} in the \textit{2}-player limit.}
    \label{fig:3rep}
\end{figure}

An introduction to \textit{GRIM*} is given in Ref.~\cite{ref2a}, where it is defined as a slight modification of the \textit{GRIM} strategy; when players play \textit{GRIM*} (this is also a \textit{deterministic} strategy), they will surely \textit{defect} in the last round of the repeated prisoner's dilemma game since there is no incentive for the players to \textit{cooperate} in the last round. Similar to \textit{GRIM}, games involving \textit{GRIM*} have a predefined number of rounds $\Bar{m}$ (i.e., a predefined discount factor $w$; \textit{deterministic} nature of the strategy). We consider one of the players (row player; $\mathfrak{P}_1$) plays the \textit{GRIM} strategy, whereas its opponent (column player; $\mathfrak{P}_2$) plays the \textit{GRIM*} strategy (see, Fig.~\ref{fig:3rep}). We also note in Fig.~\ref{fig:3rep} that both the players, $\mathfrak{P}_1$ and $\mathfrak{P}_2$, start with cooperation $C$ (from the definitions of \textit{GRIM} and \textit{GRIM*}; see Sec.~\ref{introsec}) and continues to cooperate till the \textit{penultimate} round (denoted by $\mathit{\Bar{m}-1}$). At the $\mathit{\Bar{m}}^{th}$ round, $\mathfrak{P}_2$ (who plays \textit{GRIM*}) \textit{defects} while $\mathfrak{P}_1$ (who plays \textit{GRIM}) still \textit{cooperates}.  

Now, in this case, our calculations will be much easier (since $\mathfrak{P}_2$ defects at the last round; payoffs for both players change at the $\Bar{m}^{th}$ round) if we calculate our cumulative payoffs, for both $\mathfrak{P}_1$ and $\mathfrak{P}_2$, in terms of the number of rounds $\Bar{m}$, rather than in terms of $w$ (i.e., the \textit{discount factor}). Both $\mathfrak{P}_1$ and $\mathfrak{P}_2$ starts with $C$ and continues to \textit{cooperate} (i.e., receive the \textit{reward} $\mathbb{R}$ for $\mathit{(\Bar{m}-1)}$ rounds) until the last round (i.e., the $\mathit{\Bar{m}}^{th}$ round), where $\mathfrak{P}_1$ still \textit{cooperates}, hence receiving the \textit{sucker's payoff} $\mathbb{S}$, while $\mathfrak{P}_2$ \textit{defects}, hence receiving the \textit{temptation} $\mathbb{T}$ (see, Fig.~\ref{fig:3rep}; $\mathfrak{P}_2$ is playing \textit{GRIM*}). Hence, we can write the cumulative payoffs (denoted by $\Lambda$), coupled with the number of rounds $\Bar{m}$, for both players, as \cite{ref2a, ref2b},
\begin{gather}
    \Lambda_{\mathfrak{P}_1} = (\Bar{m}-1)\mathbb{R} + \mathbb{S},\label{eq14rep}\\
    \Lambda_{\mathfrak{P}_2} = (\Bar{m}-1)\mathbb{R} + \mathbb{T}. \label{eq15rep}
\end{gather}
Similarly, if $\mathfrak{P}_1$ played \textit{GRIM*} while $\mathfrak{P}_2$ played \textit{GRIM}, then the payoffs for $\mathfrak{P}_1$ and $\mathfrak{P}_2$ given in Eqs.~(\ref{eq14rep}, \ref{eq15rep}) would just get swapped. For a very large number of rounds $\Bar{m}$, we can easily show that $\Bar{m}$ is related to the \textit{discount factor} $w$ via the relation: $\Bar{m} = \frac{1}{1-w}$ \cite{ref2b}. Hence, the results obtained for the payoffs, in Eqs.~(\ref{eq14rep}, \ref{eq15rep}), can be rewritten in terms of $w$ as,
\begin{gather}
    \Lambda_{\mathfrak{P}_1} = (\Bar{m}-1)\mathbb{R} + \mathbb{S} = \mathbb{S} + \dfrac{w\mathbb{R}}{1-w},\label{eq16rep}\\
    \Lambda_{\mathfrak{P}_2} = (\Bar{m}-1)\mathbb{R} + \mathbb{T} = \mathbb{T} + \dfrac{w\mathbb{R}}{1-w}. \label{eq17rep}
\end{gather} 

Now, instead of one playing \textit{GRIM} and the other one playing \textit{GRIM*}, if both the players play \textit{GRIM*}, it is apparent that both will receive the cumulative payoff (coupled with the \textit{discount factor} $w$),
\begin{gather}
    \Lambda_{\mathfrak{P}_1} = \Lambda_{\mathfrak{P}_2} = \mathbb{P}+ \dfrac{w\mathbb{R}}{1-w},
    \label{eq18rep}
\end{gather}
since both players will eventually \textit{defect} in the last (i.e., $\Bar{m}^{th}$) round, hence receiving the \textit{punishment} payoff $\mathbb{P}$. Similarly, if both the players play \textit{GRIM}, then both will receive the cumulative payoff, coupled with the \textit{discount factor} $w$ and rounds $\Bar{m}$, as,
\begin{gather}
    \Lambda_{\mathfrak{P}_1} = \Lambda_{\mathfrak{P}_2} =  \dfrac{\mathbb{R}}{1-w} = \Bar{m}\mathbb{R},~\text{as derived in Eq.~(\ref{eq6rep})}.
    \label{eq19rep}
\end{gather}
Hence, from Eqs.~(\ref{eq14rep}-\ref{eq19rep}), we can derive the payoff matrix $\Lambda$ corresponding to the \textit{two} players who choose either of the two strategies: \textit{GRIM} or \textit{GRIM*}, (\textit{row} player: $\mathfrak{P}_1$, \textit{column} player: $\mathfrak{P}_2$),
\begin{equation}
\small{
    \Lambda = \left[\begin{array}{c|cc} 
    	 & {\textit{GRIM}} & {\textit{GRIM*}}\\ 
    	\hline \\
        {\textit{GRIM}} & \dfrac{\mathbb{R}}{1-w}, \dfrac{\mathbb{R}}{1-w} & \mathbb{S} + \dfrac{w\mathbb{R}}{1-w}, \mathbb{T} + \dfrac{w\mathbb{R}}{1-w}\\
        \\
        {\textit{GRIM*}} & \mathbb{T} + \dfrac{w\mathbb{R}}{1-w}, \mathbb{S} + \dfrac{w\mathbb{R}}{1-w} & \mathbb{P} + \dfrac{w\mathbb{R}}{1-w}, \mathbb{P} + \dfrac{w\mathbb{R}}{1-w}
    \end{array}\right]}
    \label{eq20rep}
\end{equation}
Since $\Lambda$ is a \textit{symmetric} matrix, we can equivalently write the \textit{reduced} payoff matrix (we also denote this by $\Lambda$ for brevity), for a single player (\textit{row} player: $\mathfrak{P}_1$),
\begin{equation}
    \Lambda = \left[\begin{array}{c|c c} 
    	 & {\textit{GRIM}} & {\textit{GRIM*}}\\ 
    	\hline \\
        {\textit{GRIM}} & \dfrac{\mathbb{R}}{1-w} & \mathbb{S} + \dfrac{w\mathbb{R}}{1-w}\\
        \\
        {\textit{GRIM*}} & \mathbb{T} + \dfrac{w\mathbb{R}}{1-w} & \mathbb{P} + \dfrac{w\mathbb{R}}{1-w}
    \end{array}\right].
    \label{eq21rep}
\end{equation}
From Eq.~(\ref{eq21rep}), we find that \textit{GRIM*} becomes the \textit{strict} Nash equilibrium iff ${\Lambda}_{[GRIM*, GRIM*]}$ ($\equiv \mathbb{P}+ \frac{w\mathbb{R}}{1-w}$, in Eq.~\ref{eq21rep}) $>~{\Lambda}_{[GRIM, GRIM*]}$ ($\equiv \mathbb{S} + \frac{w\mathbb{R}}{1-w}$, in Eq.~\ref{eq21rep}), i.e., if $\mathfrak{P}_1$ unilaterally changes from \textit{GRIM} to \textit{GRIM*}, then he receives a better payoff, while the payoff received by $\mathfrak{P}_2$ remains the same. This criterion, i.e., ${\Lambda}_{[GRIM*, GRIM*]}>{\Lambda}_{[GRIM, GRIM*]}$, leads to the condition: $\mathbb{P} > \mathbb{S}$, which is always true for any prisoner's dilemma game (recall, $\mathbb{T}> \mathbb{R}>\mathbb{P}>\mathbb{S}$). Hence, we can notice that \textit{GRIM} is completely dominated by \textit{GRIM*} and \textit{GRIM*} is always the \textit{strict} Nash equilibrium (independent of the \textit{discount factor} $w$ or no. of rounds $\Bar{m}$). 

\subsubsection{\label{subwsls}WSLS vs. TFT}
\begin{figure}[!ht]
    \centering
    \includegraphics[width=1\linewidth]{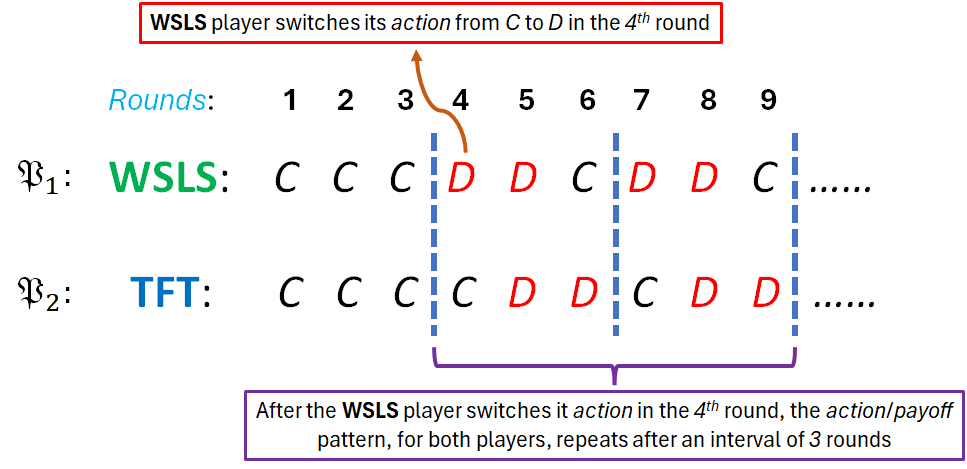}
    \caption{\textit{WSLS} vs. \textit{TFT} in the \textit{2}-player limit.}
    \label{fig:4rep}
\end{figure}

Win-stay, lose-shift (or, \textit{WSLS}) is also a \textit{deterministic} strategy that was first described by \textit{R. Axelrod} in 1984\cite{ref1}. It involves players making decisions based on the outcomes of previous rounds of the repeated prisoner's dilemma game (indicating a \textit{Markovian} nature). If a \textit{WSLS} player wins in the previous round (i.e., obtain the best feasible payoff), they stick with the same \textit{action} in the next round (hence, \textit{win-stay}), and if they lose (i.e., obtain a worse payoff), they switch to the opposite \textit{action} (hence, \textit{lose-shift}). Similar to the previous cases, games involving \textit{WSLS} have a predefined number of rounds $\Bar{m}$ (or, a predefined discount factor $w$; \textit{deterministic} nature of the strategy). Let us consider that in a \textit{2}-player repeated prisoner's dilemma game, one of the players (row player; $\mathfrak{P}_1$) plays the \textit{WSLS} strategy, whereas its opponent (column player; $\mathfrak{P}_2$) plays the \textit{TFT} strategy (see, Fig.~\ref{fig:4rep}). We note in Fig.~\ref{fig:4rep} that both the players, $\mathfrak{P}_1$ and $\mathfrak{P}_2$, start with cooperation $C$. However, in this case, we must keep in mind that $\mathfrak{P}_1$ (or, the \textit{WSLS} player) can randomly switch its action to \textit{``check"} whether it receives a better payoff against its opponent (or, $\mathfrak{P}_2$; who is playing \textit{TFT} in this case) given that its opponent still plays the same \textit{action} as it played in the previous round. A visualization of this situation is given in Fig.~\ref{fig:4rep}, where we notice that till the \textit{third} round, both $\mathfrak{P}_1$ and $\mathfrak{P}_2$ \textit{cooperate} and both receive the \textit{reward} payoff $\mathbb{R}$ (recall, $\mathbb{R}$ is not the maximum payoff one can have; \textit{temptation} $\mathbb{T}$ is the maximum payoff). However, let us say, in the \textit{fourth} round, the \textit{WSLS} player, i.e., $\mathfrak{P}_1$, switches its action to \textit{defect}, to \textit{``check"} whether it receives a better payoff than its previous round's payoff, and indeed $\mathfrak{P}_1$ receives a better payoff, i.e., the \textit{temptation} $\mathbb{T}$, given that its opponent still \textit{cooperates} in the \textit{fourth} round. 

However, since $\mathfrak{P}_2$ plays \textit{TFT}, we can easily guess that $\mathfrak{P}_2$ will switch its \textit{action} from \textit{cooperate} to \textit{defect} in the \textit{fifth} round (see, Fig.~\ref{fig:4rep}), while $\mathfrak{P}_1$ still chooses to \textit{defect} (since, it received the best possible payoff $\mathbb{T}$ in the \textit{fourth} round; hence \textit{win-stay}). In the \textit{fifth} round, both receive the \textit{punishment} payoff $\mathbb{P}$ (which is, yet again, not the best possible payoff). Hence, the \textit{WSLS} player $\mathfrak{P}_1$ again switches its action from \textit{defect} to \textit{cooperate} (i.e., \textit{lose-shift}), in the \textit{sixth} round, to \textit{``check"} whether it receives a better payoff than its previous round's payoff ($\mathbb{P}$). Since the \textit{TFT} player $\mathfrak{P}_2$ copies $\mathfrak{P}_1$'s action in the previous round, in the \textit{sixth} round, $\mathfrak{P}_2$ will \textit{defect} while $\mathfrak{P}_1$ chooses to \textit{cooperate}, indicating that $\mathfrak{P}_1$ receives the \textit{sucker's payoff} $\mathbb{S}$ while $\mathfrak{P}_2$ receives the \textit{temptation} $\mathbb{T}$ in the \textit{sixth} round. From Fig.~\ref{fig:4rep}, it can be easily verified that this pattern of ``\textit{action} selection", by both $\mathfrak{P}_1$ and $\mathfrak{P}_2$, is followed in the subsequent rounds, with a periodicity of \textit{three} rounds. 

Following Fig.~\ref{fig:4rep}, we can write the cumulative payoff (denoted by $\Lambda$), coupled with the \textit{discount factor} $w$ defined for subsequent rounds of the game (to account for the decreasing present value of future payoffs over time), for $\mathfrak{P}_1$ and $\mathfrak{P}_2$ as (see, Ref.~\cite{ref2b} for more details regarding the derivation),
\begin{gather}
    \Lambda_{\mathfrak{P}_1} = \mathbb{R} + w\mathbb{R} + w^2 \mathbb{R} + (w^3\mathbb{T} + w^4\mathbb{P} + w^5\mathbb{S} + w^6\mathbb{T} \nonumber\\
    + w^7\mathbb{P} + w^8\mathbb{S} + \ldots),\nonumber\\
    \text{or,}~\Lambda_{\mathfrak{P}_1} = \mathbb{R} + w\mathbb{R} + w^2 \mathbb{R} + w^3\Bigg[\dfrac{\mathbb{T} + w\mathbb{P} + w^2\mathbb{S}}{1-w^3}\Bigg],\label{eq22rep}\\
    \text{and,}~\Lambda_{\mathfrak{P}_2} = \mathbb{R} + w\mathbb{R} + w^2 \mathbb{R} + (w^3\mathbb{S} + w^4\mathbb{P} + w^5\mathbb{T}\nonumber\\
    + w^6\mathbb{S} + w^7\mathbb{P} + w^8\mathbb{T} + \ldots),\nonumber\\
    \text{or,}~\Lambda_{\mathfrak{P}_2} = \mathbb{R} + w\mathbb{R} + w^2 \mathbb{R} + w^3\Bigg[\dfrac{\mathbb{S} + w\mathbb{P} + w^2\mathbb{T}}{1-w^3}\Bigg].\label{eq23rep}
\end{gather}
Similarly, if $\mathfrak{P}_1$ played \textit{TFT} while $\mathfrak{P}_2$ played \textit{WSLS}, then the payoffs for $\mathfrak{P}_1$ and $\mathfrak{P}_2$ given in Eqs.~(\ref{eq22rep}, \ref{eq23rep}) would just get swapped. Now, instead of one playing \textit{WSLS} and the other one playing \textit{TFT}, we consider the case when both the players play \textit{WSLS}. In such a scenario, to make our calculations simpler, we assume that both the players, even though both are independent, initially \textit{cooperate} till the \textit{third} round and then simultaneously switch to \textit{defect} in the \textit{fourth} round, to \textit{``check"} whether they would receive a better payoff than the previous rounds.  
\begin{figure}[H]
    \centering
    \includegraphics[width=1\linewidth]{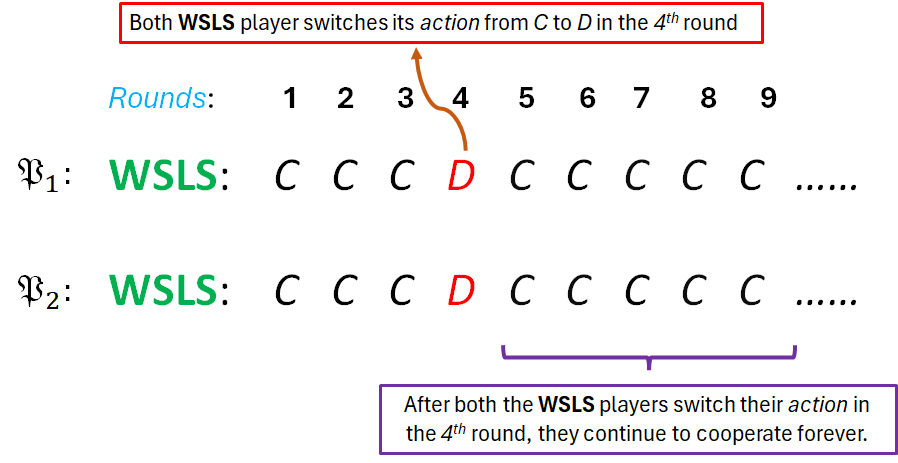}
    \caption{\centering{Both $\mathfrak{P}_1$ and $\mathfrak{P}_2$ are playing \textit{WSLS} in the \textit{2}-player repeated prisoner's dilemma.}}
    \label{fig:5rep}
\end{figure}

As evident from Fig.~\ref{fig:5rep}, in the \textit{fourth} round, both $\mathfrak{P}_1$ and $\mathfrak{P}_2$ receive the \textit{punishment} payoff $\mathbb{P}$, and hence, they both switch back to \textit{cooperate} in the \textit{fifth} round and continue to \textit{cooperate} for the rest of the game. It is apparent that both will receive the cumulative payoff (coupled with the \textit{discount factor} $w$),
\begin{gather}
    \Lambda_{\mathfrak{P}_1} = \Lambda_{\mathfrak{P}_2} = \mathbb{R}+w\mathbb{R} + w^2\mathbb{R} + w^3 \mathbb{P}+ \dfrac{w^4\mathbb{R}}{1-w}.
    \label{eq24rep}
\end{gather}
Similarly, if both the players play \textit{TFT}, then both will receive the cumulative payoff, coupled with the \textit{discount factor} $w$, as,
\begin{gather}
    \Lambda_{\mathfrak{P}_1} = \Lambda_{\mathfrak{P}_2} =  \dfrac{\mathbb{R}}{1-w} \equiv \mathbb{R} + w\mathbb{R} + w^2\mathbb{R} + \dfrac{w^3\mathbb{R}}{1-w},
    \label{eq25rep}
\end{gather}
since both players will try not to \textit{defect} as the first one to defect will receive the \textit{temptation} payoff $\mathbb{T}$ while its opponent will receive the \textit{sucker's payoff} $\mathbb{S}$ (which is clearly not a viable payoff for \textit{rational} players; both of them look for payoff maximization), and since both are playing \textit{TFT}, if any one of the players \textit{defects} in any round, then both players will receive a combination of $\mathbb{T}$ and $\mathbb{S}$ for the game's later rounds.

Hence, from Eqs.~(\ref{eq22rep}-\ref{eq25rep}), we can write the \textit{reduced} payoff matrix $\Lambda$ for a single player (\textit{row} player: $\mathfrak{P}_1$) who chooses either of the two strategies: \textit{WSLS} or \textit{TFT}, as,
\begin{equation}
\small{
    \Lambda = \left[\begin{array}{c|c | c} 
    	 & {\textit{WSLS}} & {\textit{TFT}}\\ 
    	\hline \\
        {\textit{WSLS}} & \mathbb{R}+w\mathbb{R} + w^2\mathbb{R} & \mathbb{R} + w\mathbb{R} + w^2 \mathbb{R} \\
        & + w^3 \mathbb{P}+ \dfrac{w^4\mathbb{R}}{1-w} &    + w^3\Bigg[\dfrac{\mathbb{T} + w\mathbb{P} + w^2\mathbb{S}}{1-w^3}\Bigg]     \\
        \\
        {\textit{TFT}} & \mathbb{R} + w\mathbb{R} + w^2 \mathbb{R} & \mathbb{R} + w\mathbb{R} + w^2\mathbb{R}  \\
        & + w^3\Bigg[\dfrac{\mathbb{S} + w\mathbb{P} + w^2\mathbb{T}}{1-w^3}\Bigg] & + \dfrac{w^3\mathbb{R}}{1-w}
    \end{array}\right].}
    \label{eq26rep}
\end{equation}
In the $\Lambda$ given in Eq.~(\ref{eq26rep}), we find that the term $(\mathbb{R} + w\mathbb{R} + w^2 \mathbb{R})$ is common in all the payoffs. Hence, we can remove it as it is a constant cutoff. We rewrite our payoff matrix $\Lambda$ as,
\begin{equation}
\small{
    \Lambda = \left[\begin{array}{c|c c} 
    	 & {\textit{WSLS}} & {\textit{TFT}}\\ 
    	\hline \\
        {\textit{WSLS}} & w^3 \bigg[\mathbb{P}+ \dfrac{w\mathbb{R}}{1-w}\bigg] &    w^3\Bigg[\dfrac{\mathbb{T} + w\mathbb{P} + w^2\mathbb{S}}{1-w^3}\Bigg]     \\
        \\
        {\textit{TFT}} & w^3\Bigg[\dfrac{\mathbb{S} + w\mathbb{P} + w^2\mathbb{T}}{1-w^3}\Bigg] & \dfrac{w^3\mathbb{R}}{1-w}
    \end{array}\right].}
    \label{eq27rep}
\end{equation}
Further, we can remove the multiplicative factor of $w^3$ in each term of $\Lambda$ in Eq.~(\ref{eq27rep}) as again it is a constant, and again rewrite the payoff matrix as,
\begin{equation}
{
    \Lambda = \left[\begin{array}{c|c c} 
    	 & {\textit{WSLS}} & {\textit{TFT}}\\ 
    	\hline \\
        {\textit{WSLS}} & \mathbb{P}+ \dfrac{w\mathbb{R}}{1-w} &    \dfrac{\mathbb{T} + w\mathbb{P} + w^2\mathbb{S}}{1-w^3}     \\
        \\
        {\textit{TFT}} & \dfrac{\mathbb{S} + w\mathbb{P} + w^2\mathbb{T}}{1-w^3} & \dfrac{\mathbb{R}}{1-w}
    \end{array}\right].}
    \label{eq28rep}
\end{equation}
From Eq.~(\ref{eq28rep}), we find that \textit{WSLS} becomes the \textit{strict} Nash equilibrium iff ${\Lambda}_{[WSLS, WSLS]}$ ($\equiv \mathbb{P}+ \frac{w\mathbb{R}}{1-w}$, in Eq.~\ref{eq28rep}) $>~{\Lambda}_{[TFT, WSLS]}$ ($\equiv \frac{\mathbb{S} + w\mathbb{P} + w^2\mathbb{T}}{1-w^3}$, in Eq.~\ref{eq28rep}), i.e., if $\mathfrak{P}_1$ unilaterally changes from \textit{TFT} to \textit{WSLS}, then he receives a better payoff, while the payoff received by $\mathfrak{P}_2$ remains the same. This criterion, i.e., ${\Lambda}_{[WSLS, WSLS]}>{\Lambda}_{[TFT, WSLS]}$, leads to the condition: $(\mathbb{P}-\mathbb{S}) - (\mathbb{P-R})(w-w^3) + w^2 (\mathbb{R-T})>0$, and this does not generate a closed form for the critical \textit{real} value of $w$ (see, Ref.~\cite{ref2b} for more details), above which \textit{WSLS} would become the \textit{strict} Nash equilibrium.

\subsubsection{WSLS vs. GRIM}
\begin{figure}[!ht]
    \centering
    \includegraphics[width=1\linewidth]{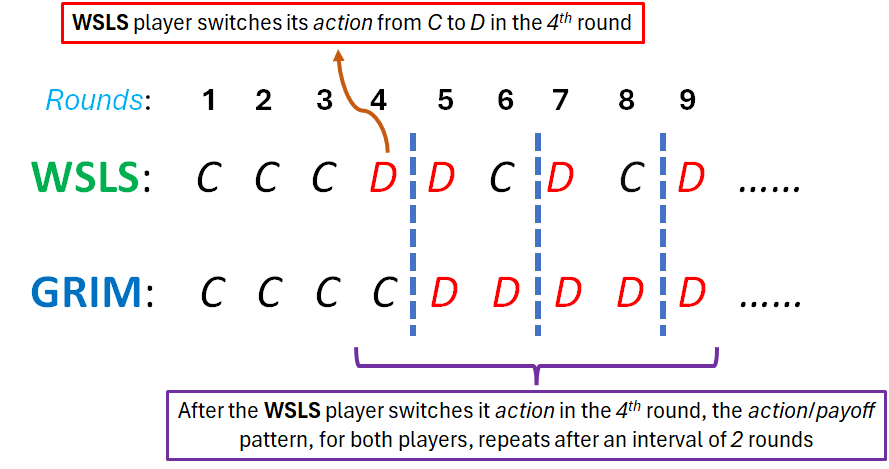}
    \caption{\textit{WSLS} vs. \textit{GRIM} in the \textit{2}-player limit.}
    \label{fig:6rep}
\end{figure}

Finally, we take into consideration that one of the players (say, the row player; $\mathfrak{P}_1$) in a \textit{2}-player repeated prisoner's dilemma game plays the \textit{WSLS} strategy, whereas its opponent (say, the column player; $\mathfrak{P}_2$) plays the \textit{GRIM} strategy (see, Fig.~\ref{fig:6rep}). We also note in Fig.~\ref{fig:6rep} that both the players, $\mathfrak{P}_1$ and $\mathfrak{P}_2$, start with cooperation $C$. However, $\mathfrak{P}_1$ (or, the \textit{WSLS} player) can randomly switch its action to \textit{``check"} whether it receives a better payoff against its opponent (or, $\mathfrak{P}_2$; who is playing \textit{GRIM} in this case) given that its opponent still plays the same \textit{action} as it played in the previous round. A visualization of this situation is given in Fig.~\ref{fig:6rep}, where we notice that till the \textit{third} round, both $\mathfrak{P}_1$ and $\mathfrak{P}_2$ \textit{cooperate} and both receive the \textit{reward} payoff $\mathbb{R}$ (recall, $\mathbb{R}$ is not the maximum payoff one can have; \textit{temptation} $\mathbb{T}$ is the maximum payoff). However, in the \textit{fourth} round, the \textit{WSLS} player, i.e., $\mathfrak{P}_1$, switches its action to \textit{defect}, to \textit{``check"} whether it receives a better payoff than its previous round's payoff, and indeed $\mathfrak{P}_1$ receives a better payoff, i.e., the \textit{temptation} $\mathbb{T}$, given that its opponent still \textit{cooperates} in the \textit{fourth} round. 

Since $\mathfrak{P}_2$ plays \textit{GRIM}, we can easily guess that $\mathfrak{P}_2$ will permanently switch its \textit{action} from \textit{cooperate} to \textit{defect} from the \textit{fifth} round (see, Fig.~\ref{fig:6rep}). Meanwhile, $\mathfrak{P}_1$ also chooses to \textit{defect} in the \textit{fifth} round (since it received the best possible payoff $\mathbb{T}$ in the \textit{fourth} round; hence \textit{win-stay}) resulting in both receiving the \textit{punishment} payoff $\mathbb{P}$ (which is, yet again, not the best possible payoff). Hence, the \textit{WSLS} player $\mathfrak{P}_1$ again switches its action from \textit{defect} to \textit{cooperate} (i.e., \textit{lose-shift}), in the \textit{sixth} round, to \textit{``check"} whether it receives a better payoff than its previous round's payoff ($\mathbb{P}$). Since the \textit{GRIM} player $\mathfrak{P}_2$ has permanently shifted to \textit{defect}, from Fig.~\ref{fig:6rep}, it can be easily verified that a pattern of both $(\mathfrak{P_1}:~defect,~\mathfrak{P}_2:~defect)$ and $(\mathfrak{P_1}:~cooperate,~\mathfrak{P}_2:~defect)$ is followed in the subsequent rounds, with a periodicity of \textit{two} rounds. 

Following Fig.~\ref{fig:6rep}, we can write the cumulative payoffs (denoted by $\Lambda$), coupled with the \textit{discount factor} $w$ (to account for the decreasing present value of future payoffs over time), for both players as\cite{ref2b},
\begin{gather}
    \Lambda_{\mathfrak{P}_1} = \mathbb{R} + w\mathbb{R} + w^2 \mathbb{R} + w^3\mathbb{T} + (w^4\mathbb{P} + w^5\mathbb{S} + w^6\mathbb{P}  \nonumber\\
    + w^7\mathbb{S} + w^8\mathbb{P} + \ldots),\nonumber\\
    \text{or,}~\Lambda_{\mathfrak{P}_1} = \mathbb{R} + w\mathbb{R} + w^2 \mathbb{R} + w^3\mathbb{T} + w^4\Bigg[\dfrac{\mathbb{P} + w\mathbb{S}}{1-w^2}\Bigg],\label{eq29rep}\\
    \text{and,}~\Lambda_{\mathfrak{P}_2} = \mathbb{R} + w\mathbb{R} + w^2 \mathbb{R} + w^3\mathbb{S} + (w^4\mathbb{P} + w^5\mathbb{T} \nonumber\\
    + w^6\mathbb{P} + w^7\mathbb{T} + w^8\mathbb{P} + \ldots),\nonumber\\
    \text{or,}~\Lambda_{\mathfrak{P}_2} = \mathbb{R} + w\mathbb{R} + w^2 \mathbb{R} + w^3\mathbb{S} + w^4\Bigg[\dfrac{\mathbb{P} + w\mathbb{T}}{1-w^2}\Bigg].\label{eq30rep}
\end{gather}
Similarly, if $\mathfrak{P}_1$ played \textit{GRIM} while $\mathfrak{P}_2$ played \textit{WSLS}, then the payoffs for $\mathfrak{P}_1$ and $\mathfrak{P}_2$ given in Eqs.~(\ref{eq29rep}, \ref{eq30rep}) would just get swapped. Now, instead of one playing \textit{WSLS} and the other one playing \textit{GRIM}, we consider the case when both the players play \textit{WSLS}. In such a scenario, we follow the exact formalism as described in the Subsection~\ref{subwsls} (see, Eq.~(\ref{eq24rep})), and we have the cumulative payoff (coupled with the \textit{discount factor} $w$) as,
\begin{gather}
    \Lambda_{\mathfrak{P}_1} = \Lambda_{\mathfrak{P}_2} = \mathbb{R}+w\mathbb{R} + w^2\mathbb{R} + w^3 \mathbb{P}+ \dfrac{w^4\mathbb{R}}{1-w}.
    \label{eq31rep}
\end{gather}
Similarly, if both the players play \textit{GRIM}, then both will receive the cumulative payoff, coupled with the \textit{discount factor} $w$, as (see, Eq.~(\ref{eq19rep})),
\begin{gather}
    \Lambda_{\mathfrak{P}_1} = \Lambda_{\mathfrak{P}_2} =  \dfrac{\mathbb{R}}{1-w},
    \label{eq32rep}
\end{gather}
Hence, following the same formalism as we discussed in the \textit{WSLS} vs. \textit{TFT} case, from Eqs.~(\ref{eq29rep}-\ref{eq32rep}), we can derive the \textit{reduced} payoff matrix $\Lambda$ corresponding to a single player (\textit{row} player: $\mathfrak{P}_1$) who chooses either of the two strategies: \textit{WSLS} or \textit{GRIM}, as,
\begin{equation}
{
    \Lambda = \left[\begin{array}{c|c c} 
    	 & {\textit{WSLS}} & {\textit{GRIM}}\\ 
    	\hline \\
        {\textit{WSLS}} & \mathbb{P}+ \dfrac{w\mathbb{R}}{1-w} &    \mathbb{T} + w\Bigg[\dfrac{\mathbb{P} + w\mathbb{S}}{1-w^2}\Bigg]     \\
        \\
        {\textit{GRIM}} & \mathbb{S} + w\Bigg[\dfrac{\mathbb{P} + w\mathbb{T}}{1-w^2}\Bigg] & \dfrac{\mathbb{R}}{1-w}
    \end{array}\right].}
    \label{eq33rep}
\end{equation}
From Eq.~(\ref{eq33rep}), we find that \textit{WSLS} becomes the \textit{strict} Nash equilibrium iff ${\Lambda}_{[WSLS, WSLS]}$ ($\equiv \mathbb{P}+ \frac{w\mathbb{R}}{1-w}$, in Eq.~\ref{eq33rep}) $>~{\Lambda}_{[GRIM, WSLS]}$ ($\equiv \mathbb{S} + w[\frac{\mathbb{P} + w\mathbb{T}}{1-w^2}]$, in Eq.~\ref{eq33rep}), i.e., if $\mathfrak{P}_1$ unilaterally changes from \textit{GRIM} to \textit{WSLS}, then he receives a better payoff, while the payoff received by $\mathfrak{P}_2$ remains the same. This criterion, i.e., ${\Lambda}_{[WSLS, WSLS]}>{\Lambda}_{[GRIM, WSLS]}$, leads to the condition: $\mathbb{P}(1-w-w^2) - \mathbb{R}w(1+w) - \mathbb{S}(1-w^2) + \mathbb{T}w^2 >0$, and this does not generate a closed form for the critical \textit{real} value of $w$ (see, Ref.~\cite{ref2b} for more details), above which \textit{WSLS} would become the \textit{strict} Nash equilibrium.  

In this section, we discussed how the players would behave when they play the \textit{repeated} prisoner's dilemma game, in the \textit{2}-player setting, by considering \textit{five} strategy sets, namely, \textit{(TFT~\text{vs.}~All-D)}, \textit{(GRIM {vs.}~All-D)}, \textit{(GRIM~\text{vs.}~GRIM*)}, \textit{(WSLS~\text{vs.}~TFT)} and \textit{(WSLS~\text{vs.}~GRIM)}. However, in our work, we aim to study the players' behaviour in the \textit{thermodynamic} (or, \textit{infinite} player) limit. In the next section, we will discuss the analytical \textit{Nash equilibrium mapping} (NEM) and numerical \textit{Agent-based modelling} (ABM) techniques that, in the thermodynamic limit, we shall employ to examine these games. This discussion will be followed by a detailed analysis of the results obtained for the games using these two techniques.

\section{\label{therm}Thermodynamic limit of Repeated games}
\subsection{Motivation: Why do we study games in the thermodynamic limit?}
Game theoretic models, such as the Prisoner's dilemma, offer insights into human behaviour within populations \cite{ref1,ref2a}, including larger groups. As the number of players increases (i.e., approaches the \textit{thermodynamic} limit), cooperation can become more advantageous due to repeated interactions and reputation effects, offering an evolutionary context for strategy evolution \cite{ref1}. Studying games in this limit helps uncover the dynamics of \textit{cooperation}, \textit{competition}, and \textit{strategy evolution} within large populations, providing deeper insights into real-world strategic behaviour. This motivates our exploration of repeated games in the thermodynamic limit, aiming to understand the emergence of cooperative behaviour among an infinite number of players across different game scenarios.

In this paper, we try to understand the emergence of cooperative behaviour among an \textit{infinite} number of players playing the \textit{five} different types of repeated Prisoner's dilemma (RPD) game, with the help of analytical and numerical techniques based on the \textit{1D}-Ising chain. To study the behaviour of players, we establish a mapping between different types of RPD games and the \textit{1D}-Ising chain and use the game analogue of the thermodynamic magnetization, i.e., game magnetization, to analyze the player's behaviour.

\subsection{Introduction to the two methods}
\subsubsection{\label{NEM}Nash equilibrium mapping (NEM)}
In NEM, we analytically map a social dilemma game to a \textit{spin-1/2} $1D$-Ising chain (i.e., the \textit{thermodynamic} limit) (see also Refs.~\cite{ref6, ref10, ref12}). We consider a $2$-strategy; $2$-player social dilemma while mapping it to a $2$-site Ising chain. The 2 strategies (say, $\$_1$ and $\$_2$) have a \textit{one-to-one} mapping to the \textit{2}-spin (say, $\pm1$) $1D$-Ising chain and we have the $2$-site $(\text{say,}~A~\text{and}~B)$ $1D$-Ising chain Hamiltonian as \cite{ref12},
\begin{equation}
    H = -\mathcal{T}(\mathfrak{\sigma}_A\mathfrak{\sigma}_B + \mathfrak{\sigma}_B\mathfrak{\sigma}_A) - \mathcal{F}(\mathfrak{\sigma}_A + \mathfrak{\sigma}_B) = \Delta_A + \Delta_B,
    \label{eq2.12rep}
\end{equation}
where, $\mathcal{T}$ is the coupling constant, $\mathcal{F}$ is the external field, $\mathfrak{\sigma}_i ~\forall~i\in \{A,B\},$ is the spin (either $+1$ or $-1$) at the $i^{th}$ site, and $\Delta_i$ denotes the energy of the $i^{th}$ site. Each of the individual site's energy is given as,
\begin{equation}
    \Delta_A = -\mathcal{T}\mathfrak{\sigma}_A\mathfrak{\sigma}_B - \mathcal{F}\mathfrak{\sigma}_A,~\text{and}~\Delta_B = -\mathcal{T}\mathfrak{\sigma}_B\mathfrak{\sigma}_A - \mathcal{F}\mathfrak{\sigma}_B.
    \label{eq2.13}
\end{equation}
Here, the total \textit{two}-spin Ising chain energy: $\Delta= \Delta_A +\Delta_B$. In social dilemmas, the maximization of the player's feasible payoffs corresponds to finding the Nash equilibrium of the game. However, when we consider a $1D$-Ising chain, we look to minimize $\Delta$ in order to reach the energy equilibrium condition. Hence, we equate the social dilemma payoff matrix to the negative of the energy matrix in order to establish a link between the Nash equilibrium of the social dilemma game and the Ising chain's energy equilibrium configuration \cite{ref6}. Each element of $\Delta$ (i.e., $\Delta_i$) corresponds to a particular pair of spin values $(\sigma_A,\sigma_B)$, i.e., $(\sigma_A,\sigma_B)\in \{(\pm 1, \pm 1)\}$, at each $1D$-Ising chain site since game payoff maximization indicates negative energy minimization. Thus,
\begin{equation}
\begin{footnotesize}
    -\Delta = 
    \left[\begin{array}{c|c c} 
    	 & \mathfrak{\sigma}_B = +1 & \mathfrak{\sigma}_B = -1\\ 
    	\hline 
    	\mathfrak{\sigma}_A = +1 & (\mathcal{T}+\mathcal{F}), (\mathcal{T}+\mathcal{F}) & (\mathcal{F}-\mathcal{T}), -(\mathcal{F}+\mathcal{T})\\
        \mathfrak{\sigma}_A = -1 & -(\mathcal{F}+\mathcal{T}), (\mathcal{F}-\mathcal{T}) & (\mathcal{T}-\mathcal{F}), (\mathcal{T}-\mathcal{F})
    \end{array}\right].
    \label{eq2.14rep}
\end{footnotesize} 
\end{equation}
For a general $2$-player; $2$-strategy (say, $\$_1$ and $\$_2$) \textit{symmetric} social dilemma game, we have the social dilemma payoff matrix $\Lambda'$ as,
\begin{equation}
    \Lambda = \left[\begin{array}{c|c c} 
    	 & \$_1 & \$_2\\ 
    	\hline 
    	\$_1 & \mathrm{m, m} & \mathrm{n, p}\\
        \$_2 & \mathrm{p, n} & \mathrm{q, q}
    \end{array}\right]. 
    \label{eq2.15rep}
\end{equation}
where, $(\mathrm{m,n,p, q})$ are defined as the social dilemma payoffs. Making use of a series of linear transformations on $\Lambda'$ in Eq.~(\ref{eq2.15rep}), to establish a \textit{one-to-one} correspondence between the payoffs in Eq.~(\ref{eq2.15rep}) and the energy matrix $\Delta$ of the two-spin Ising chain in Eq.~(\ref{eq2.14rep}), that preserves the Nash equilibrium (see, Ref.~\cite{ref6} and \textbf{Appendix} of Ref.~\cite{ref5a} for detailed calculations),
\begin{equation}
    \mathrm{m}\rightarrow \frac{\mathrm{m}-\mathrm{p}}{2},~\mathrm{n} \rightarrow \frac{\mathrm{n}-\mathrm{q}}{2},~\mathrm{p}\rightarrow \frac{\mathrm{p}-\mathrm{m}}{2},~\mathrm{q}\rightarrow \frac{\mathrm{q}-\mathrm{n}}{2},
    \label{eq2.16rep}
\end{equation}
we equate $\Lambda'$ (see, Eq.~(\ref{eq2.15rep})) to $-\Delta$ (see, Eq.~(\ref{eq2.14rep})), to rewrite the Ising parameters $(\mathcal{T},\mathcal{F})$ in terms of the social dilemma payoffs as \cite{ref6, ref10, ref12},
\begin{equation}
    \mathcal{F} = \frac{\mathrm{(m-p)+(n-q)}}{4},~ \text{and}~\mathcal{T} = \frac{\mathrm{(m-p)-(n-q)}}{4}.
    \label{eq2.17rep}
\end{equation}
In $1D$-Ising chains, the parameter $\beta$ is defined as proportional to the temperature ($T$) inverse, or, $\beta = \frac{1}{k_B T}$, where $k_B$ is the \textit{Boltzmann constant}, and in game theoretic models, temperature is a proxy for \textit{selection pressure} in player's strategy selection, and this is also termed \textit{noise}. Thus, $\beta \rightarrow \infty$ (or, $T\rightarrow 0$) implies \textit{zero noise}, meaning that the players' strategies remain unchanged, whereas, $\beta \rightarrow 0$ (or, $T\rightarrow \infty$) implies \textit{infinite noise}, meaning that the players' strategy selection is entirely random. We can also interpret $\beta$ as the \textit{selection intensity} (see, Ref.~\cite{ref13}) where, for $\beta\ll 1$, we observe strategies being randomly selected, whereas, for $\beta\gg 1$, we find \textit{zero} randomness in strategy selection. 

Since we consider the \textit{game magnetization} as our game's order parameter, to determine the game magnetization $\mu$, we need to derive the NEM partition function using the game payoffs as a basis. For the given $H$ (see, Eq.~(\ref{eq2.12rep})) and Eq.~(\ref{eq2.17rep}), the partition function $\Upsilon^{NEM}$ can be written in terms of the payoffs $(\mathrm{m, n, p, q})$ as \cite{ref12},
\begin{gather}
    \Upsilon^{Ising} = \sum_{\{\sigma_i\}} e^{-\beta H} = e^{2\beta (\mathcal{T}+\mathcal{F})} + e^{2\beta (\mathcal{T}-\mathcal{F})} + 2e^{-2\beta\mathcal{T}},\nonumber\\
    \text{or,}~\Upsilon^{NEM} = e^{\beta\mathrm{(m-p)}} + e^{-\beta \mathrm{(n+q)}} + 2e^{\frac{\beta}{2} \mathrm{(n+p-m-q)}},
    \label{eq2.18}
\end{gather}
where, $\beta = \frac{1}{k_B T}$ indicates the \textit{noise/selection pressure}. 

\textbf{Game magnetization:} We obtain the game magnetization ($\mu^{NEM}$) in relation to the game payoffs by utilising the payoff matrix $\Lambda$ defined in Eq.~(\ref{eq2.15rep}) and partition function $\Upsilon^{NEM}$, see also \cite{ref12}, 
\begin{equation}
    \mu^{NEM} = \dfrac{1}{\beta}\dfrac{\partial}{\partial \mathcal{F}}\ln{\Upsilon^{NEM}} =\dfrac{\sinh{\beta \frac{\mathrm{(m-p)+(n-q)}}{4}}}{\mathfrak{Z}},
    \label{eq22}    
\end{equation}
with, $\mathfrak{Z} = \sqrt{\sinh^2{\beta \frac{\mathrm{(m-p)+(n-q)}}{4}} + e^{-4\beta \frac{\mathrm{(m-p)-(n-q)}}{4}}}$. The payoffs $(\mathrm{m,n,p,q})$ represent the payoffs associated with the \textit{five} different types of RPD that were discussed in Sec.~\ref{2rept} (see, the payoff matrices $\Lambda$ derived in Eqs.~(\ref{eq13rep}, \ref{eq7rep}, \ref{eq21rep}, \ref{eq28rep}, \ref{eq33rep})).

\subsubsection{\label{ABM}Agent-based Modelling (ABM)}
ABM \cite{ref10, ref12} is a numerical modelling technique that has been used to analyze \textit{one-shot} games in the \textit{thermodynamic} limit. However, to the best of our knowledge, ABM has not been previously implemented to study the emergence of cooperative behaviour in \textit{repeated} social dilemmas like the repeated Prisoner's dilemma game. Hence, the main attraction of our work is that we numerically analyze the emergence of cooperation among an infinite number of players, playing the \textit{repeated} Prisoner's dilemma game. We take into consideration \textit{1000} players who reside on the $1D$-Ising chain, and they interact with their nearest neighbours only in the presence of a periodic boundary condition. The energy matrix $\Delta$ is just the negative of the payoff matrix $\Lambda$, as defined for the \textit{five} cases, and this gives the Ising chain's individual site energy. At this point, we modify the player's strategy by iterating through a conditional loop \textit{1,000,000} times, which amounts to an average of \textit{1000} strategy modifications per player. Herein below is a synopsis of the ABM algorithm:
\begin{enumerate}
    \item At each $1D$-Ising chain site, assign a random strategy to each player: \textit{0} (say, for strategy $\$_1$) or \textit{1} (say, for strategy $\$_2$). \textbf{NOTE}: we can synonymously use the words \textit{``action"} and \textit{``strategy"} for \textit{one-shot} games; for \textit{repeated} games, we assign a random \textit{strategy} (not \textit{action}) to each player, at each Ising site and the \textit{strategy} is based on a series of \textit{action} adopted by the player.
    \item Choose a principal player at random to ascertain both its unique strategy as well as its closest neighbour's strategy. The energy $\Delta$ of the principal player is ascertained based on the strategies that have been determined. The principal player's energy is computed for each of the two scenarios: either it chose the opposite strategy while preserving the closest neighbour's strategy, or it chose the same strategy as its closest neighbour. 
    \item For each of the two possible outcomes, the energy difference ($d\Delta$) is determined for the principal player. The principal player's current strategy is inverted depending on whether the Fermi function $(1 + e^{\beta \cdot d\Delta})^{-1} > 0.5$; if not, it is not flipped~\cite{ref21}. 
    \item Now, to determine the \textit{game magnetization} $\mu^{ABM}$, after each run of the conditional spin-flipping loop, we calculate the difference between the number of players playing strategy $\$_1$ and the number of players playing strategy $\$_2$. This gives the total magnetization $\Tilde{\mu}= \sum_{i} \sigma_i$, for $\sigma_i$ being the strategy ($0$ or $1$) of the player at the $i^{th}$-site, for each cycle of the conditional loop. Finally, we take the average of the total magnetization for all the loops to determine the overall game magnetization, i.e., $\mu^{ABM} = \langle \Tilde{\mu}  \rangle$.
    \item Proceed to step 2 and carry out this procedure a \textit{1000} times.
\end{enumerate}
After reviewing the Python code uploaded in \href{https://github.com/rajdeep2810/RCPD---Agent-based-modelling}{GitHub} \cite{ref26}, one can have a better understanding of this algorithm. Given that our primary goal is to maximise the feasible payoff — which we can only do when our system reaches the energy equilibrium or the least energy configuration — we see that the likelihood of strategy switching drops as the energy difference $d\Delta$ increases. Refs.~\cite{ref12, ref12a} also provide a detailed explanation of the algorithm's basic structure (which is based on the \textit{Metropolis algorithm} \cite{ref21, ref22}).

\subsection{Results and Analysis}
\begin{figure*}[!ht]
    \centering
    \begin{subfigure}[b]{\columnwidth}
        \centering
        \includegraphics[width = 0.9\textwidth]{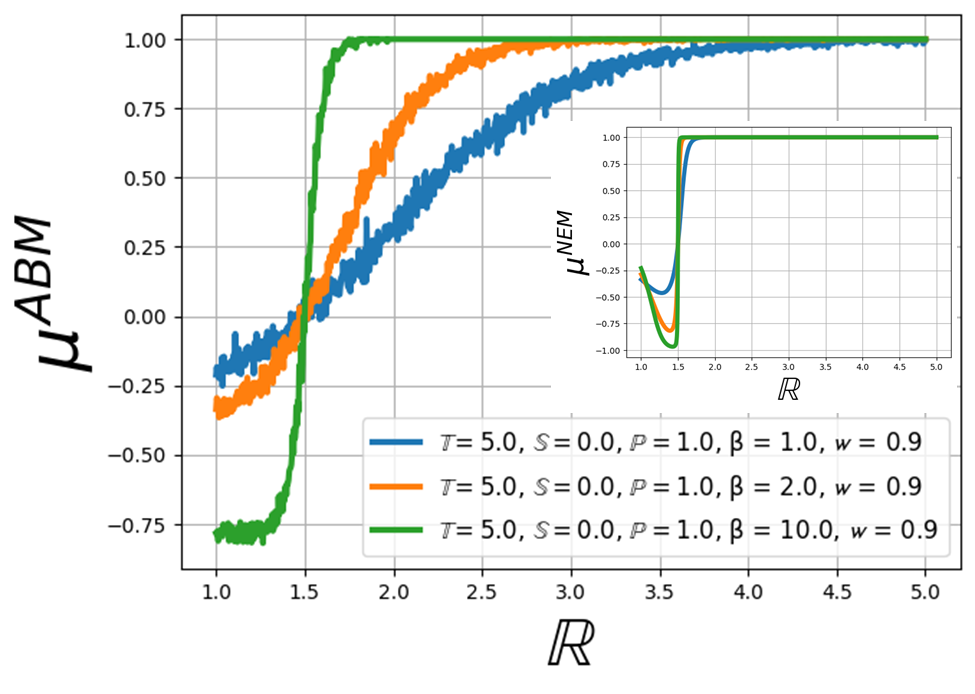}
        \caption{$\mu^{ABM/NEM}$ vs. $\mathbb{R}$}
        \label{fig7a}
    \end{subfigure}
    \begin{subfigure}[b]{\columnwidth}
        \centering
        \includegraphics[width = 0.9\textwidth]{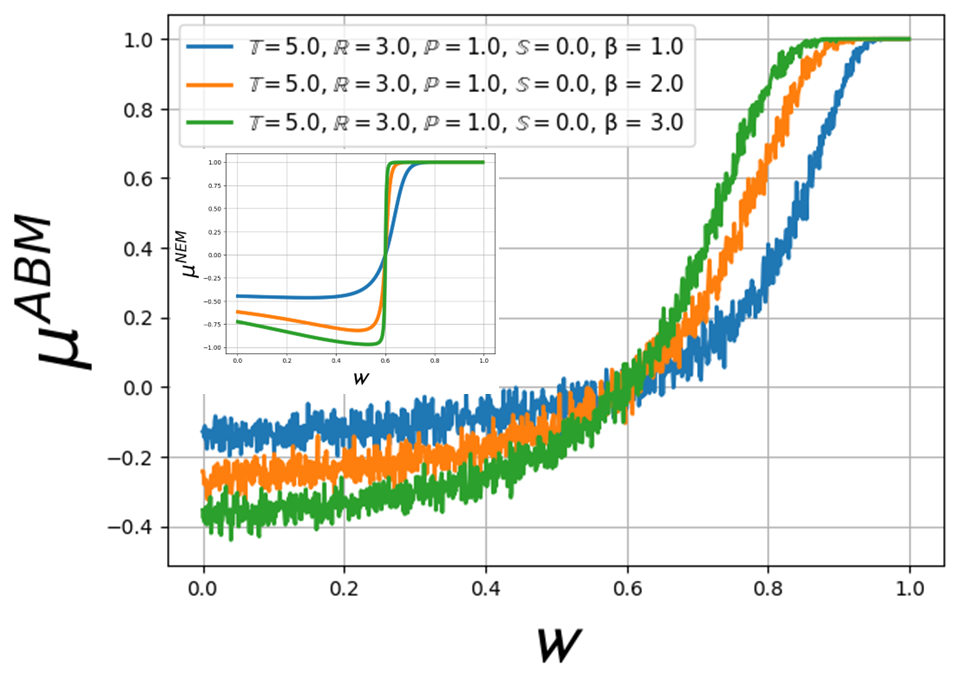}
        \caption{$\mu^{ABM/NEM}$ vs. $w$}
        \label{fig7b}
    \end{subfigure}
    \caption{\centering{ABM and NEM (in \textit{insets}) for both \textit{GRIM} vs. \textit{All-D} and \textit{TFT} vs. \textit{All-D}: (\ref{fig7a}) \textbf{Game magnetization} $\mu$ vs. \textbf{reward} $\mathbb{R}$ for $w=0.9$, i.e., \textit{10 rounds}, and (\ref{fig7b}) $\mu$ vs. \textbf{discount factor} $w$ for $\mathbb{R}=3.0$. In both cases, \textbf{sucker's payoff} $\mathbb{S} = 0.0$, \textbf{temptation} $\mathbb{T} = 5.0$, and \textbf{punishment} $\mathbb{P}= 1.0$, respectively. The phase transition occurs at: (\ref{fig7a}) $\mathbb{R}_c = 1.4$ and (\ref{fig7b}) $w_c = 0.6$.}}
    \label{fig:7rep}
\end{figure*} 

\subsubsection{GRIM vs. All-D (or, TFT vs. All-D)}
For both cases of \textit{GRIM} vs. \textit{All-D} and \textit{TFT} vs. \textit{All-D}, by comparing Eqs.~(\ref{eq13rep}, \ref{eq7rep}, \ref{eq2.15rep}), we have $\mathrm{m} = \frac{\mathbb{R}}{1-w}$, $\mathrm{n} = \mathbb{S} + \frac{w\mathbb{P}}{1-w}$, $\mathrm{p} = \mathbb{T} + \frac{w\mathbb{P}}{1-w}$ and $\mathrm{q} = \frac{\mathbb{P}}{1-w}$, respectively. Putting these values in Eq.~(\ref{eq2.17rep}) will give us the Ising parameters $(\mathcal{T},\mathcal{F})$ in terms of the payoffs (defined for this case) as,
\begin{gather}
    \mathcal{T} = \dfrac{1}{4}\bigg[\dfrac{\mathbb{R} + (1-2w)\mathbb{P}}{1-w} - \mathbb{T} - \mathbb{S} \bigg],\nonumber\\
    \text{and}~\mathcal{F} = \dfrac{1}{4}\bigg[\dfrac{\mathbb{R} - \mathbb{P}}{1-w} - \mathbb{T} + \mathbb{S} \bigg].
    \label{eqfein}
\end{gather}
The NEM game magnetization $\mu^{NEM}$ can be easily calculated by putting the expressions of $(\mathcal{T},\mathcal{F})$, given in Eq.~(\ref{eqfein}), in Eq.~(\ref{eq22}), i.e.,
\begin{equation}
    \mu^{NEM} =\dfrac{\sinh{\beta \mathcal{F} }}{\sqrt{\sinh^2{\beta \mathcal{F}} + e^{-4\beta \mathcal{T}}}}.
    \label{eqbetatft}
\end{equation}
The ABM game magnetization $\mu^{ABM}$ can be determined following the algorithm described in Sec.~\ref{ABM}. We then plot $\mu^{ABM}$ and $\mu^{NEM}$ against a changing \textit{reward} $\mathbb{R}$ and discount factor $w$ in Fig.~\ref{fig:7rep}.

From Fig.~\ref{fig7a}, we notice that, in the thermodynamic limit, for increasing $\mathbb{R}$, a greater proportion of players tend to shift from \textit{All-D} to \textit{GRIM/TFT} (i.e., $\mu^{ABM/NEM}\rightarrow +1$) since \textit{GRIM/TFT} becomes the strict Nash equilibrium (NE) strategy in this scenario. For the given $w = 0.9$, $\mathbb{T} = 5.0$, $\mathbb{P} = 1.0$ and $\mathbb{S} = 0.0$, we find that for \textit{GRIM/TFT} to be the strict NE, $\mathbb{R}>1.4$ (from the criterion for \textit{GRIM/TFT} to be strict NE: $w > \frac{\mathbb{T-R}}{\mathbb{T-P}}$). In both ABM and NEM plots, we find that for $\mathbb{R}>1.4$, a majority of players switch to \textit{GRIM/TFT}, i.e., this becomes the dominant strategy of the population, and the transition point is independent of \textit{noise} $\beta$. Similarly, in Fig.~\ref{fig7b}, for increasing $w$, we observe a change in NE strategy from \textit{All-D} to \textit{GRIM/TFT} (i.e., $\mu^{ABM/NEM}\rightarrow +1$). The number of rounds is given by $\Bar{m} = 1/(1-w)$, and for both NEM and ABM, we observe that \textit{GRIM/TFT} becomes the dominant strategy of the players when $w>0.6$. This indicates that for \textit{GRIM/TFT} to become the strict NE of the repeated game, we must have a minimum of $\Bar{m} = 1/(1-w) = $ $1/(1-0.6) = \lfloor 2.5\rfloor = 3$ rounds in the \textit{repeated} prisoner's dilemma game (independent of noise $\beta$), i.e., the phase transition occurs at $\bar{m}_c \simeq 3$. The minimum number of rounds for \textit{GRIM/TFT} to become the strict NE of \textit{repeated} prisoner's dilemma is \textit{3} since in these particular cases, if players playing \textit{GRIM} or \textit{TFT} encounters \textit{defection} in the first round, it responds with \textit{defection} in the second round. However, it's only in or after the third round that it can potentially win over the payoff lost in the \textit{first} round (in the \textit{first} round, the \textit{GRIM/TFT} player gets the sucker's payoff $\mathbb{S}$) due to its opponent defecting in that round.

\begin{figure}
    \centering
    \includegraphics[width=0.9\linewidth]{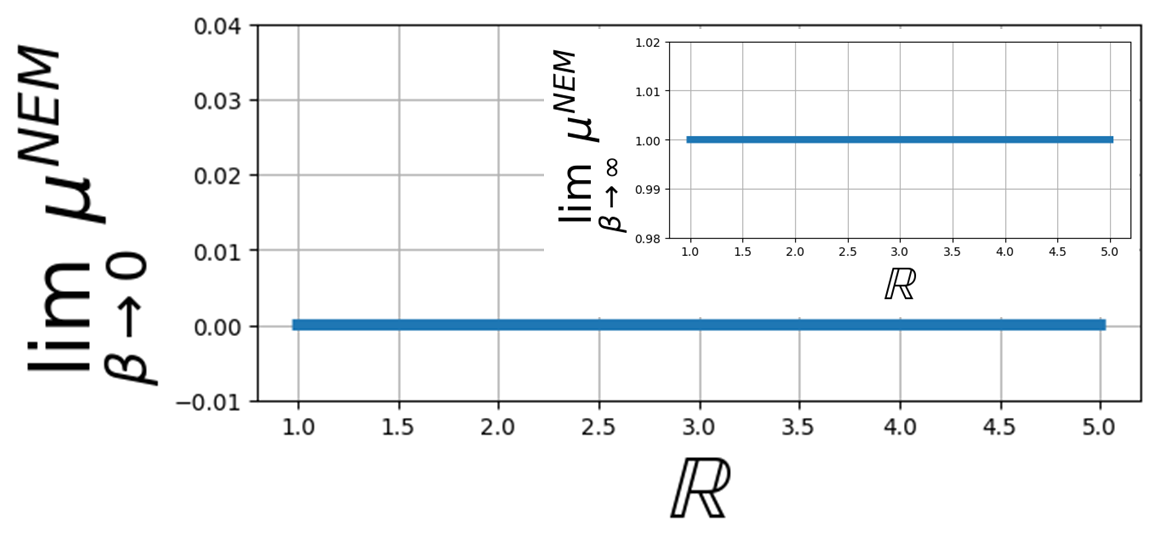}
    \caption{NEM: \textbf{Game magnetization} $\mu$ vs. \textbf{reward} $\mathbb{R}$ in the \textit{infinite noise} (i.e., $\beta\rightarrow 0$) and \textit{zero noise} (i.e., $\beta\rightarrow \infty$; in \textit{insets}) limits. In both cases, \textbf{discount factor} $w = 0.99$, \textbf{sucker's payoff} $\mathbb{S} = 0.0$, \textbf{temptation} $\mathbb{T} = 5.0$, and \textbf{punishment} $\mathbb{P}= 1.0$, respectively}
    \label{fig:lim1}
\end{figure}

What is interesting is that in the \textit{2}-player repeated prisoner's dilemma game, where the players are playing either \textit{GRIM} (or, \textit{TFT}) or \textit{All-D}, the critical values of $\mathbb{R}_c$ and $w_c$ match with the values obtained in the \textit{infinite}-player repeated prisoner's dilemma game. However, the nature of both these games is significantly different. In the \textit{2}-player game, the phase transition (at $\mathbb{R}_c$ and $w_c$) is marked by a \textit{discontinuity} (where \textit{2} players change their strategy abruptly when $w>\frac{\mathbb{T-R}}{\mathbb{T-P}}$), whereas in the \textit{infinite}-player game, the phase transition is \textit{smooth} and the phase transition depends on the \textit{noise} $\beta$ (see Fig.~\ref{fig:7rep}). In both Figs.~\ref{fig7a}, \ref{fig7b}, in the \textit{zero noise} (i.e., $\beta\rightarrow\infty$) limit, we observe a \textit{first}-order phase transition among the players from $\textit{All-D}\rightarrow \textit{GRIM/TFT}$ at $w = w_c = \frac{\mathbb{T-R}}{\mathbb{T-P}}$, i.e., $\mu^{ABM/NEM}: -1\rightarrow +1$. For increasing $w$ (say $w=0.99$, i.e., number of \textit{rounds} $\bar{m} = 1/(1-w) = 100$), as shown in the \textit{inset} of Fig.~\ref{fig:lim1}, we find that all the players shift from $\textit{All-D}\rightarrow \textit{GRIM/TFT}$ even at very low values of \textit{reward} $\mathbb{R}$. This indicates that for an increasing \textit{discount factor} $w$ (i.e., increasing number of \textit{rounds} $\bar{m}$), the rate of shift from $\textit{All-D}\rightarrow \textit{GRIM/TFT}$ also increases among the players. 

Similarly, in the \textit{infinite noise} (i.e., $\beta\rightarrow 0$) limit (see, Figs.~\ref{fig:7rep}, \ref{fig:lim1}), we find $\mu^{ABM/NEM}\rightarrow 0$ indicating players' arbitrary choice of strategies, i.e., equiprobable selection of \textit{All-D} and \textit{GRIM} (or, \textit{TFT}).

\subsubsection{GRIM vs. GRIM*}
\begin{figure}[!ht]
    \centering
    \includegraphics[width = 0.9\columnwidth]{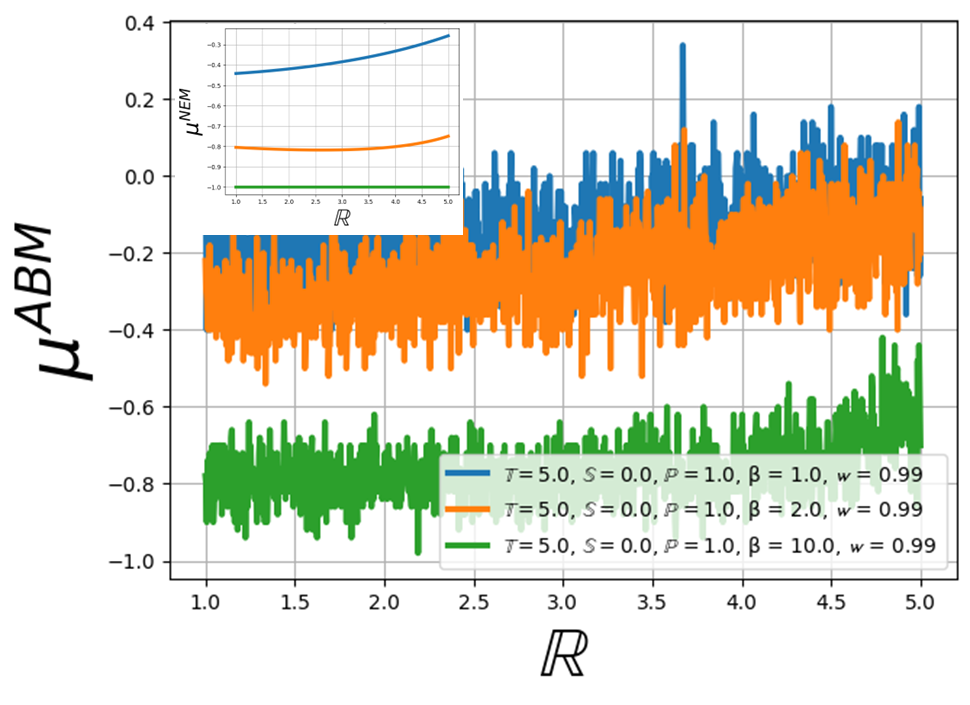}
    \caption{\centering{\centering{ABM and NEM (in \textit{inset}) for \textit{GRIM} vs. \textit{GRIM*}: \textbf{Game magnetization} $\mu$ vs. \textbf{reward} $\mathbb{R}$ for fixed $w=0.99$ (i.e., $\bar{m}=\frac{1}{1-w} = 100$ rounds), \textbf{temptation} $\mathbb{T}=5.0$, \textbf{sucker's payoff} $\mathbb{S} = 0.0$ and \textbf{punishment} $\mathbb{P} = 1.0$.}}}
    \label{fig:9rep}
\end{figure}

For the case of \textit{GRIM} vs. \textit{GRIM*} case, by comparing Eqs.~(\ref{eq21rep}) and (\ref{eq2.15rep}), we have $\mathrm{m} = \frac{\mathbb{R}}{1-w}$, $\mathrm{n} = \mathbb{S} + \frac{w\mathbb{R}}{1-w}$, $\mathrm{p} = \mathbb{T} + \frac{w\mathbb{R}}{1-w}$ and $\mathrm{q} = \mathbb{P} + \frac{w\mathbb{R}}{1-w}$, respectively. Putting these values in Eq.~(\ref{eq2.17rep}) will give us the Ising parameters $(\mathcal{T},\mathcal{F})$ in terms of the payoffs (defined for this case) as,
\begin{gather}
    \mathcal{T} = \dfrac{1}{4}[\mathbb{R} +\mathbb{P} - \mathbb{T} - \mathbb{S}],~\text{and}~\mathcal{F} = \dfrac{1}{4}[\mathbb{R} + \mathbb{S} - \mathbb{T} - \mathbb{P}].
    \label{eqfein1}
\end{gather}
The NEM game magnetization $\mu^{NEM}$ can be easily calculated by putting the expressions of $(\mathcal{T},\mathcal{F})$, given in Eq.~(\ref{eqfein1}), in Eq.~(\ref{eq22}), i.e.,
\begin{equation}
    \mu^{NEM} =\dfrac{\sinh{\beta \mathcal{F} }}{\sqrt{\sinh^2{\beta \mathcal{F}} + e^{-4\beta \mathcal{T}}}}.
\end{equation}
The ABM game magnetization $\mu^{ABM}$ can be determined following the algorithm described in Sec.~\ref{ABM}. We then plot $\mu^{ABM}$ and $\mu^{NEM}$ against a changing \textit{reward} $\mathbb{R}$ in Fig.~\ref{fig:9rep}.

\begin{figure}
    \centering
    \includegraphics[width=0.9\linewidth]{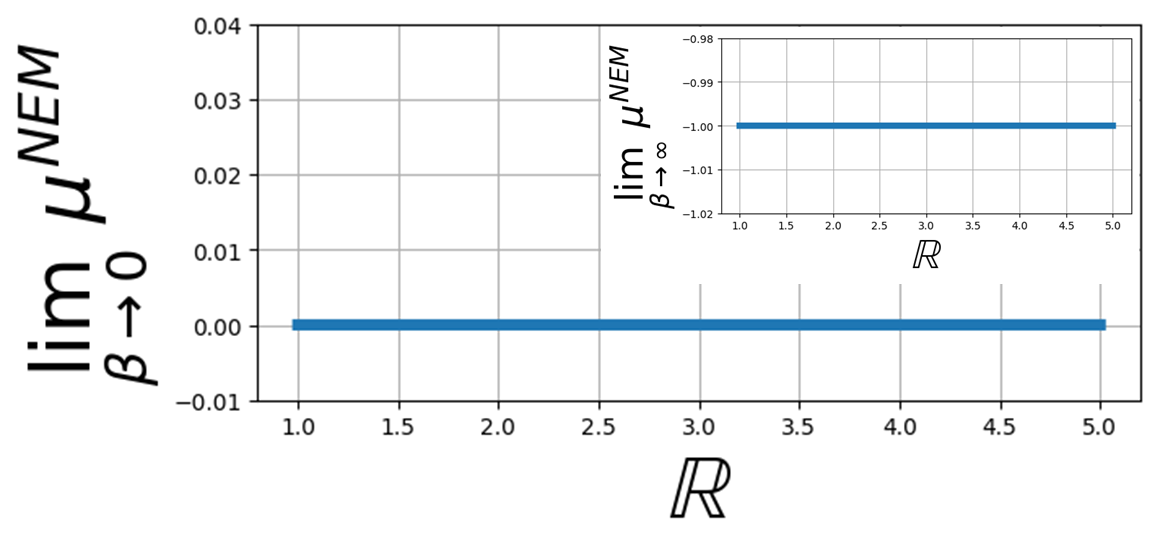}
    \caption{NEM: \textbf{Game magnetization} $\mu$ vs. \textbf{reward} $\mathbb{R}$ in the \textit{infinite noise} (i.e., $\beta\rightarrow 0$) and \textit{zero noise} (i.e., $\beta\rightarrow \infty$; in \textit{insets}) limits. In both cases, \textbf{discount factor} $w = 0.99$ (i.e., $\bar{m}=\frac{1}{1-w} = 100$ rounds), \textbf{sucker's payoff} $\mathbb{S} = 0.0$, \textbf{temptation} $\mathbb{T} = 5.0$, and \textbf{punishment} $\mathbb{P}= 1.0$, respectively.}
    \label{fig:lim2}
\end{figure}

In \textit{GRIM} vs. \textit{GRIM*} case, we have the Ising parameters in relation to the game payoffs: $\mathcal{T} = \frac{1}{4}[\mathbb{R+P-S-T}]$ and $\mathcal{F} = \frac{1}{4}[\mathbb{R+S-T-P}]$, where we find that both $\mathcal{T}$ and $\mathcal{F}$ are independent of $w$ (hence we do not check the variation of game magnetization with respect to $w$). The discount factor $w$ (or, the number of rounds $\Bar{m} = 1/(1-w)$) will have no effect on the player's behaviour while playing \textit{GRIM*} against \textit{GRIM}; the player's behaviour is same in both \textit{one-shot} and \textit{repeated} game settings, i.e., they always prefer to choose \textit{GRIM*} over \textit{GRIM} in both game settings indicating the absence of a phase transition in this type of RPD. For $\mathbb{T}= 5.0$, $\mathbb{P} = 1.0$, $\mathbb{S} = 0.0$ and $w=0.0$ (or, $\Bar{m}=10$ rounds), from Fig.~\ref{fig:9rep}, we notice that, in the thermodynamic limit, for increasing $\mathbb{R}$, a greater proportion of players continues to play \textit{GRIM*} over \textit{GRIM}, and \textit{GRIM*} becomes the strict Nash equilibrium (NE) strategy in this scenario (the game magnetization $\mu$ is always $-ve$ indicating players are playing \textit{GRIM*}). In the \textit{2}-player limit, we expect both players to play \textit{GRIM*} (since it is the Nash equilibrium). However, in the \textit{infinite}-player limit, we find that a greater proportion of players play the \textit{GRIM*} strategy, but not all. Moreover, what is interesting is that the noise $\beta$ plays a significant role in determining the fraction of players who choose to play \textit{GRIM*} in the thermodynamic limit (there is no such \textit{noise} dependency in the \textit{2}-player limit). As shown in the \textit{inset} of Fig.~\ref{fig:lim2}, in the \textit{zero noise} (i.e., $\beta\rightarrow\infty$) limit, $\mu^{ABM/NEM} \rightarrow -1$ indicating that all players are playing \textit{GRIM*}. Similarly, in the \textit{infinite noise} (i.e., $\beta\rightarrow 0$) limit (see, Fig.~\ref{fig:lim2}), we find $\mu^{ABM/NEM}\rightarrow 0$ indicating players' arbitrary choice of strategies, i.e., equiprobable selection of \textit{GRIM*} and \textit{GRIM} strategies.

\subsubsection{WSLS vs. TFT}
\begin{figure}[!ht]
    \centering
    \includegraphics[width = 0.9\columnwidth]{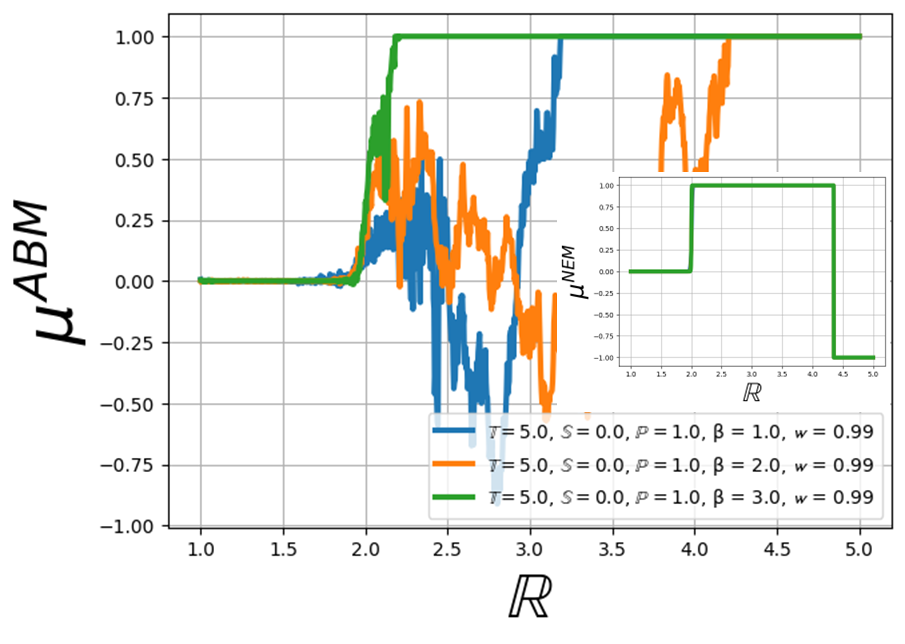}
    \caption{\centering{\centering{ABM and NEM (in \textit{inset}) for \textit{WSLS} vs. \textit{TFT}: \textbf{Game magnetization} $\mu$ vs. \textbf{reward} $\mathbb{R}$ for fixed $w=0.99$, \textbf{temptation} $\mathbb{T}=5.0$, \textbf{sucker's payoff} $\mathbb{S} = 0.0$ and \textbf{punishment} $\mathbb{P} = 1.0$. The phase transition occurs at $\mathbb{R}_c \approx2.0$.}}}
    \label{fig:10rep}
\end{figure}

For the case of \textit{WSLS} vs. \textit{TFT} case, by comparing Eqs.~(\ref{eq28rep}) and (\ref{eq2.15rep}), we have $\mathrm{m} = \mathbb{P}+ \frac{w\mathbb{R}}{1-w}$, $\mathrm{n} = \frac{\mathbb{T} + w\mathbb{P} + w^2\mathbb{S}}{1-w^3}$, $\mathrm{p} = \frac{\mathbb{S} + w\mathbb{P} + w^2\mathbb{T}}{1-w^3}$ and $\mathrm{q} = \frac{\mathbb{R}}{1-w}$, respectively. Putting these values in Eq.~(\ref{eq2.17rep}) will give us the Ising parameters $(\mathcal{T},\mathcal{F})$ in terms of the payoffs (defined for this case) as,
\begin{gather}
    \mathcal{T} = \dfrac{1}{4}\bigg[ \bigg( \dfrac{1-2w - w^3}{1-w^3} \bigg)\mathbb{P}  + \bigg(\dfrac{1+w}{1-w} \bigg)\mathbb{R} - \bigg(\dfrac{1+w^2}{1-w^3} \bigg)\nonumber\\ (\mathbb{S} + \mathbb{T}) \bigg],~\mathcal{F} = \dfrac{1}{4}\bigg[\mathbb{P} - \mathbb{R} + \bigg( \dfrac{1-w^2}{1-w^3}\bigg)(\mathbb{T} - \mathbb{S}) \bigg].
    \label{eqfein2}
\end{gather}
The NEM game magnetization $\mu^{NEM}$ can be easily calculated by putting the expressions of $(\mathcal{T},\mathcal{F})$, given in Eq.~(\ref{eqfein2}), in Eq.~(\ref{eq22}), i.e.,
\begin{equation}
    \mu^{NEM} =\dfrac{\sinh{\beta \mathcal{F} }}{\sqrt{\sinh^2{\beta \mathcal{F}} + e^{-4\beta \mathcal{T}}}}.
\end{equation}
The ABM game magnetization $\mu^{ABM}$ can be determined following the algorithm described in Sec.~\ref{ABM}. We then plot $\mu^{ABM}$ and $\mu^{NEM}$ against a changing \textit{reward} $\mathbb{R}$ in Fig.~\ref{fig:10rep}.

We see in Fig.~\ref{fig:10rep} that a greater proportion of players tends to shift to \textit{WSLS} in the thermodynamic limit for increasing $\mathbb{R}$ (i.e., $\mu^{ABM/NEM}\rightarrow +1$), indicating that \textit{WSLS} becomes the strict Nash equilibrium (NE) strategy in this scenario. Since we cannot determine the exact value of $w$ (hence, $\mathbb{R}$; the payoffs and $w$ are related to each other) where this phase transition occurs (see, Sec.~\ref{subwsls} for discussion), our only resort is to analyze the results obtained for NEM and ABM and approximately determine the payoff $\mathbb{R}$ where we see the phase transition. From Fig.~\ref{fig:10rep}, we can see that at $\mathbb{R}_c\approx 2.0$, the phase transition occurs. Previously, for $\mathbb{R}<2.0$, players would equiprobably choose either \textit{WSLS} or \textit{TFT} (owing to low \textit{reward} associated with the game), i.e., $\mu^{NEM/ABM}\rightarrow 0$. However, for $\mathbb{R}>2.0$, all the players shift to \textit{WSLS} because, for increasing \textit{reward} $\mathbb{R}$, \textit{WSLS} capitalizes on opportunities to continue exploiting \textit{cooperative} opponents after a win, leading to potentially higher payoffs in environments with larger \textit{rewards} (like \textit{temptation} $\mathbb{T}$). 

\begin{figure}
    \centering
    \includegraphics[width=0.9\linewidth]{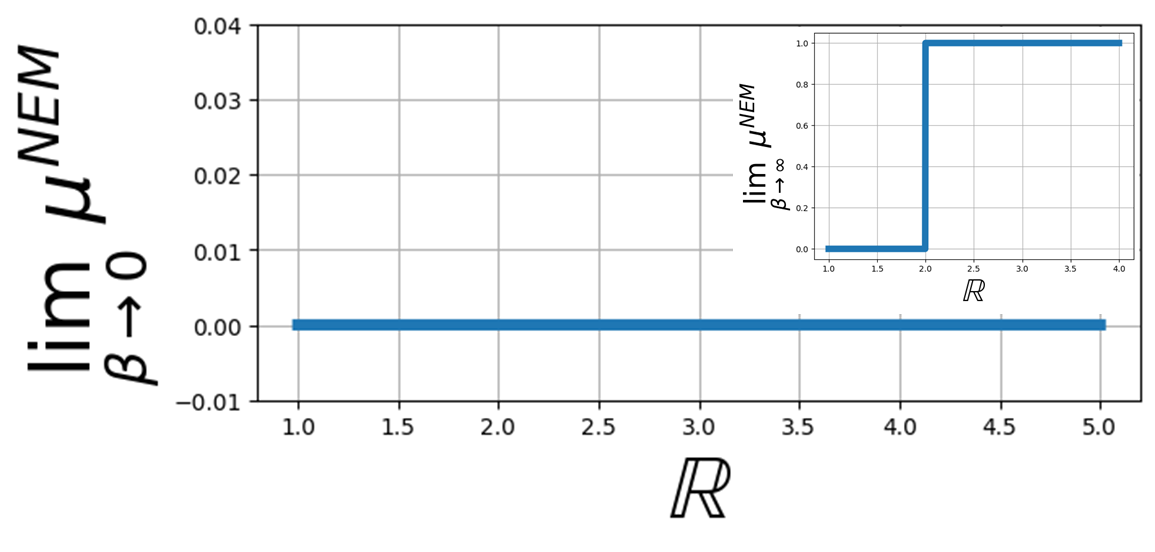}
    \caption{NEM: \textbf{Game magnetization} $\mu$ vs. \textbf{reward} $\mathbb{R}$ in the \textit{infinite noise} (i.e., $\beta\rightarrow 0$) and \textit{zero noise} (i.e., $\beta\rightarrow \infty$; in \textit{insets}) limits. In both cases, \textbf{discount factor} $w = 0.99$, \textbf{sucker's payoff} $\mathbb{S} = 0.0$, \textbf{temptation} $\mathbb{T} = 5.0$, and \textbf{punishment} $\mathbb{P}= 1.0$, respectively}
    \label{fig:lim3}
\end{figure}

Readers might have noticed in Fig.~\ref{fig:10rep} that for $\mathbb{R}>4.0$, $\mu^{NEM}\rightarrow -1$ whereas $\mu^{ABM}\rightarrow +1$. A possible explanation for this is the assumptions that we have considered while defining the \textit{WSLS} vs. \textit{TFT} case, such as the \textit{WSLS} player can only change his decision independently once (at the \textit{fourth} round); after that, its behaviour is governed by the action of the \textit{TFT} player (see, Fig.~\ref{fig:4rep}). Also, this anomaly in the NEM and ABM results for $\mathbb{R}>4.0$ might be due to a lower number of loops being considered for the ABM algorithm. We plan to check the results for ten million loops and compare them with the NEM results, but this is computationally expensive. As shown in the \textit{inset} of Fig.~\ref{fig:lim3}, in the \textit{zero noise} (i.e., $\beta\rightarrow\infty$) limit, $\mu^{ABM/NEM} \rightarrow 1,~\forall~\mathbb{R}>2.0$ indicating that all players are playing \textit{WSLS} when the \textit{reward} $\mathbb{R}>2.0$. However, in the same \textit{zero noise} limit, $\mu^{ABM/NEM} \rightarrow 0,~\forall~\mathbb{R}<2.0$ indicating random strategy selection by the players when the \textit{reward} $\mathbb{R}<2.0$ (i.e., a low \textit{reward}). Similarly, in the \textit{infinite noise} (i.e., $\beta\rightarrow 0$) limit (see, Fig.~\ref{fig:lim3}), we find $\mu^{ABM/NEM}\rightarrow 0$ indicating players' arbitrary choice of strategies, i.e., equiprobable selection of \textit{WSLS} and \textit{TFT} strategies. 

\subsubsection{\label{wsgr}WSLS vs. GRIM}
\begin{figure}
    \centering
    \includegraphics[width = 0.9\columnwidth]{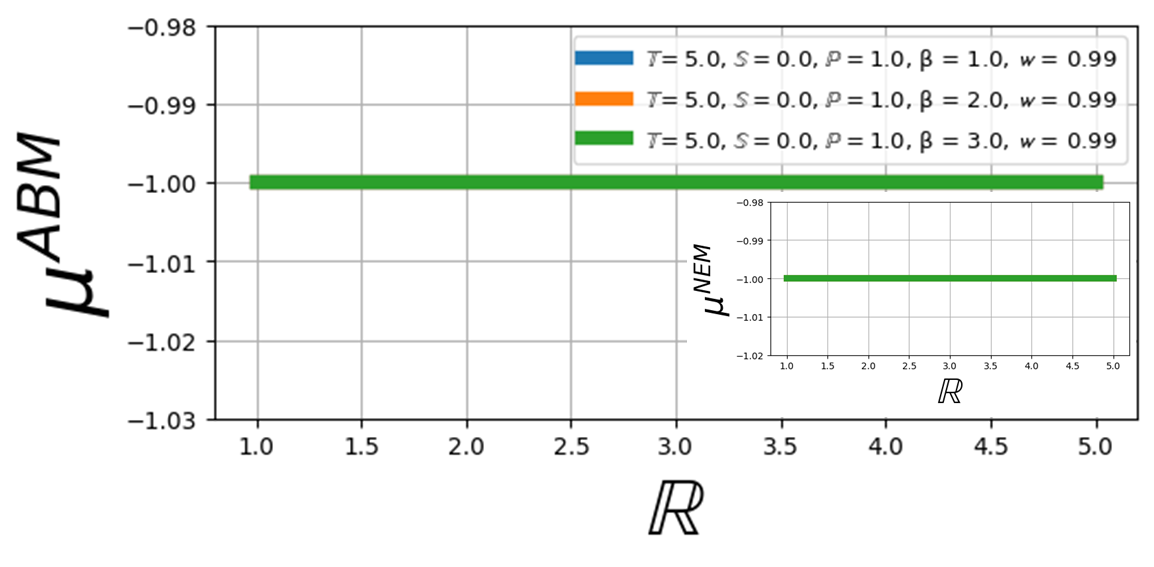}
    \caption{\centering{\centering{ABM and NEM (in \textit{inset}) for \textit{WSLS} vs. \textit{GRIM}: \textbf{Game magnetization} $\mu$ vs. \textbf{reward} $\mathbb{R}$ for fixed $w=0.99$, \textbf{temptation} $\mathbb{T}=5.0$, \textbf{sucker's payoff} $\mathbb{S} = 0.0$ and \textbf{punishment} $\mathbb{P} = 1.0$.}}}
    \label{fig:11rep}
\end{figure}

For the case of \textit{WSLS} vs. \textit{GRIM} case, by comparing Eqs.~(\ref{eq33rep}) and (\ref{eq2.15rep}), we have $\mathrm{m} = \mathbb{P}+ \frac{w\mathbb{R}}{1-w}$, $\mathrm{n} = \mathbb{T} + w[\frac{\mathbb{P} + w\mathbb{S}}{1-w^2}]$, $\mathrm{p} = \mathbb{S} + w[\frac{\mathbb{P} + w\mathbb{T}}{1-w^2}]$ and $\mathrm{q} = \frac{\mathbb{R}}{1-w}$, respectively. Putting these values in Eq.~(\ref{eq2.17rep}) will give us the Ising parameters $(\mathcal{T},\mathcal{F})$ in terms of the payoffs (defined for this case) as,
\begin{gather}
    \mathcal{T} = \dfrac{1}{4}\bigg[ \bigg( \dfrac{1-2w - w^2}{1-w^2} \bigg)\mathbb{P}  + \bigg(\dfrac{1+w}{1-w} \bigg)\mathbb{R} - \bigg(\dfrac{1}{1-w^2} \bigg)\nonumber\\ (\mathbb{S} + \mathbb{T}) \bigg],~\mathcal{F} = \dfrac{1}{4}\bigg[\mathbb{P} - \mathbb{R} + \bigg( \dfrac{1-2w^2}{1-w^2}\bigg)(\mathbb{T} - \mathbb{S}) \bigg].
    \label{eqfein3}
\end{gather}
The NEM game magnetization $\mu^{NEM}$ can be easily calculated by putting the expressions of $(\mathcal{T},\mathcal{F})$, given in Eq.~(\ref{eqfein3}), in Eq.~(\ref{eq22}), i.e.,
\begin{equation}
    \mu^{NEM} =\dfrac{\sinh{\beta \mathcal{F} }}{\sqrt{\sinh^2{\beta \mathcal{F}} + e^{-4\beta \mathcal{T}}}}.
\end{equation}
The ABM game magnetization $\mu^{ABM}$ can be determined following the algorithm described in Sec.~\ref{ABM}. We then plot $\mu^{ABM}$ and $\mu^{NEM}$ against a changing \textit{reward} $\mathbb{R}$ in Fig.~\ref{fig:11rep}.
In Fig.~\ref{fig:11rep}, we notice that, in the thermodynamic limit, all the players shift to \textit{GRIM} (i.e., $\mu^{ABM/NEM}\rightarrow -1,~\forall~\mathbb{R}$) irrespective of the \textit{reward} $\mathbb{R}$ and in the presence of finite \textit{noise} $\beta$ (i.e., $\beta \nrightarrow 0$), indicating that \textit{GRIM} becomes the strict Nash equilibrium (NE) strategy in this scenario. This particular case, where players have to play either \textit{WSLS} or \textit{GRIM}, leads to the same result in both \textit{2}-player and \textit{infinite}-player limits. Here, we do not see any phase transition among the players, since all the players choose \textit{GRIM} over \textit{WSLS}. 
\begin{figure}
    \centering
    \includegraphics[width=0.9\linewidth]{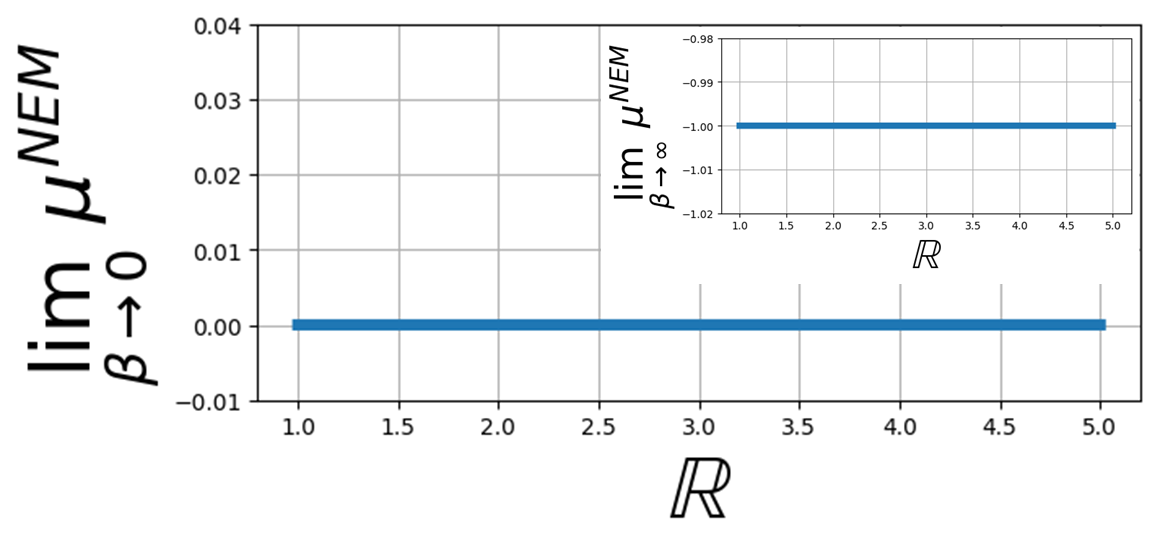}
    \caption{NEM: \textbf{Game magnetization} $\mu$ vs. \textbf{reward} $\mathbb{R}$ in the \textit{infinite noise} (i.e., $\beta\rightarrow 0$) and \textit{zero noise} (i.e., $\beta\rightarrow \infty$; in \textit{insets}) limits. In both cases, \textbf{discount factor} $w = 0.99$, \textbf{sucker's payoff} $\mathbb{S} = 0.0$, \textbf{temptation} $\mathbb{T} = 5.0$, and \textbf{punishment} $\mathbb{P}= 1.0$, respectively}
    \label{fig:lim4}
\end{figure}
As shown in the \textit{inset} of Fig.~\ref{fig:lim4}, in the \textit{zero noise} (i.e., $\beta\rightarrow\infty$) limit, $\mu^{ABM/NEM} \rightarrow -1,~\forall~\mathbb{R}$ indicating that all players are playing \textit{GRIM}. Similarly, in the \textit{infinite noise} (i.e., $\beta\rightarrow 0$) limit (see, Fig.~\ref{fig:lim4}), we find $\mu^{ABM/NEM}\rightarrow 0$ indicating players' arbitrary choice of strategies, i.e., equiprobable selection of \textit{WSLS} and \textit{GRIM} strategies. 

\begin{table*}
\centering
\renewcommand{\arraystretch}{1.75}
\resizebox{\textwidth}{!}{%
\begin{tabular}{|cc|c|c|c|c|c|}
\hline
\multicolumn{2}{|c|}{\textbf{NEM/ABM}} & \begin{tabular}[c]{@{}c@{}}\textit{GRIM} vs. \textit{All-D} \\ (or, \textit{TFT} vs. \textit{All-D}) \end{tabular} & \multicolumn{1}{c|}{\textit{GRIM*} vs. \textit{GRIM}} & \multicolumn{1}{c|}{\textit{WSLS} vs. \textit{TFT}} & \multicolumn{1}{c|}{\textit{WSLS} vs. \textit{GRIM}}\\ \hline
\multicolumn{1}{|c|}{\multirow{2}{*}{\Large$\mathbf{\mu}$}} & \multicolumn{1}{c|}{$\beta \rightarrow 0$} & \multicolumn{1}{c|}{0, $\forall~w,~\mathbb{R}$}  & \multicolumn{1}{c|}{0, $\forall~\mathbb{R}$}&  \multicolumn{1}{c|}{0, $\forall~\mathbb{R}$}  &   \multicolumn{1}{c|}{0, $\forall~w,~\mathbb{R}$}  \\ \cline{2-6} 
\multicolumn{1}{|c|}{} & \multicolumn{1}{c|}{$\beta \rightarrow \infty$} & \begin{tabular}[c]{@{}c@{}}$+ 1$, $\forall~w> \frac{\mathbb{T-R}}{\mathbb{T-P}}$\\ $-1$, $\forall~w< \frac{\mathbb{T-R}}{\mathbb{T-P}}$\end{tabular} & \multicolumn{1}{c|}{-1, $\forall~\mathbb{R}$} &  \begin{tabular}[c]{@{}c@{}}$+ 1$, $\forall~\mathbb{R}>2.0$\\ $0$, $\forall~\mathbb{R}< 2.0$\end{tabular}   &   \multicolumn{1}{c|}{-1, $\forall~w \neq \dfrac{(\mathbb{P+R}) \pm \sqrt{(\mathbb{P+R})^2 - 4(\mathbb{P-S})(\mathbb{T+S-P-R})}}{2(\mathbb{T+S-P-R})} \geq 0$} \\ \hline
\end{tabular}%
}
\caption{\centering{\textit{Five} cases of \textit{repeated} prisoner's dilemma game with  \textbf{temptation} $\mathbb{T}= 5.0$, \textbf{reward} $\mathbb{R}\in [1.0, 5.0]$, \textbf{sucker's payoff} $\mathbb{S}= 0.0$, \textbf{punishment} $\mathbb{P}= 1.0$, discount factor $w$, and measure of noise $\beta$. \textbf{NOTE:} \textit{GRIM*} vs. \textit{GRIM} is independent of $w$. For \textit{WSLS} vs. \textit{GRIM}, if $\mathbb{T} = 5.0$, $\mathbb{R} = 3.0$, $\mathbb{P} = 1.0$ and $\mathbb{S} = 0.0$, then we have $\lim_{\beta\rightarrow\infty} \mu = -1,~\forall~w \neq (2\pm\sqrt{3}) \approx 3.73~\text{or}~0.27$. As discussed earlier in Sec.~\ref{df}, we have to satisfy $w\leq 1$ and this gives us the condition: $\lim_{\beta\rightarrow\infty} \mu = -1,~\forall~w \neq 0.27$ for $\mathbb{T} = 5.0$, $\mathbb{R} = 3.0$, $\mathbb{P} = 1.0$ and $\mathbb{S} = 0.0$.}}
\label{tab:table-rcpd}
\end{table*}

\section{\label{conc-rep}Conclusion}
In this paper, we sought to understand the emergence of cooperative behaviour among an infinite number of players playing the \textit{repeated} Prisoner's dilemma (RPD) by comparing a numerical technique, i.e., Agent-based modelling (ABM), with the analytical NEM method. We considered the \textit{game magnetization} as the order parameter and used it to study the players' behaviour in the thermodynamic limit. We considered \textit{five} different types of RPD designed using \textit{five} different strategies, namely, \textit{All-D}, \textit{GRIM}, \textit{GRIM*}, \textit{TFT}, and \textit{WSLS}. For the cases of \textit{GRIM} vs. \textit{All-D}, \textit{TFT} vs. \textit{All-D}, and \textit{WSLS} vs. \textit{TFT}, we observed a \textit{first}-order phase transition among the players for increasing reward $\mathbb{R}$. Also, the phase transition has an intrinsic dependence on the \textit{noise} $\beta$ present in the game setup. This also showcases the fact that for certain cases of \textit{repeated} prisoner's dilemma, at finite noise $\beta$, we observe a change in the Nash equilibrium from one strategy to another, and this is marked by a \textit{first}-order phase transition in the game magnetization.
For these particular cases, in the \textit{2}-player limit, the phase transition point (for both $\mathbb{R}$ and $w$) is marked by a \textit{discontinuity} (where \textit{2} players change their strategy abruptly), whereas in the \textit{infinite}-player game, the phase transition is \textit{smooth} and the phase transition depends on the \textit{noise} $\beta$. Our study demonstrates a significant change in the player's behaviour when repeated games are played in the thermodynamic limit rather than in the \textit{2}-player limit.

For the remaining two cases of \textit{GRIM} vs. \textit{GRIM*} and \textit{WSLS} vs. \textit{GRIM}, we observed no such phase transitions among the players. However, in the case of \textit{GRIM} vs. \textit{GRIM*}, we found that the \textit{noise} $\beta$ influenced the fraction of players choosing \textit{GRIM*} over \textit{GRIM}. For increasing $\beta$ (or, \textit{reduced} noise/selection pressure), a greater proportion of players opted for the \textit{GRIM*} strategy over the \textit{GRIM} strategy. Table~\ref{tab:table-rcpd} summarizes the outcomes for game magnetization corresponding to each RPD type for the limiting $\beta$ values. For \textit{GRIM} vs. \textit{All-D} and \textit{TFT} vs. \textit{All-D}, we obtained exactly the same results, owing to one of the players always choosing to \textit{defect} (see Figs.~\ref{fig:2rep}, \ref{fig:1rep}), leading to the same cumulative payoff for the player playing either \textit{TFT} or \textit{GRIM}. To conclude, this paper is primarily focused on mapping a \textit{repeated} prisoner's dilemma game to the $1D$-Ising chain and then numerically as well as analytically studying the emergence of cooperative behaviour among an infinite number of players by involving game magnetization as an order parameter. We plan to extend the Nash equilibrium mapping technique to the $2D$-Ising model (and other precisely solvable statistical frameworks) or to \textit{regular} (or, \textit{complete}) graph structures, and study whether it would convey more information regarding the cooperative behaviour among an \textit{infinite} number of players playing a \textit{repeated} game. This work can also be further extended to \textit{repeated} quantum prisoner's dilemma games.


\newpage


\begin{thebibliography}{H}

\bibitem{ref1} M.A. Nowak, S. Coakley, \emph{Evolution, Games, and God: The Principle of Cooperation} (2013)

\bibitem{ref2}M. Nowak, A. Sasaki, C. Taylor \textit{et al}, \emph{Emergence of cooperation and evolutionary stability in finite populations}, \href{https://doi.org/10.1038/nature02414}{Nature \textbf{428}, 646–650 (2004).}

\bibitem{ref2a}M. Nowak, \href{https://www.hup.harvard.edu/books/9780674023383}{Evolutionary Dynamics: Exploring the Equations of Life (2006)}.

\bibitem{ref2b}A. Magazinnik, \href{https://ocw.mit.edu/courses/17-810-game-theory-spring-2021/mit17_810s21_lec5.pdf}{MIT Lecture notes - Lecture 5: Repeated Games (2021)}.

\bibitem{ref3}C. Benjamin, A. Dash, \emph{Thermodynamic susceptibility as a measure of cooperative behaviour in social dilemmas}, \href{https://doi.org/10.1063/5.0015655}{Chaos, \textbf{30} 093117 (2020).}

\bibitem{ref4}Jabir T. M., N. Vyas, C. Benjamin, \emph{Is the essence of a quantum game captured completely in the original classical game?}, \href{https://doi.org/10.1016/j.physa.2021.126360}{Physica A: Stat. Mech. and App., \textbf{584}, 126360 (2021)}

\bibitem{ref5}A. Mukhopadhyay \textit{et. al.}, \emph{Repeated quantum game as a stochastic game: Effects of the shadow of the future and entanglement}, \href{https://doi.org/10.1016/j.physa.2024.129613}{Physica A: Stat. Mech. and App., \textbf{637}, 129613 (2024)}

\bibitem{ref5a}S. Sarkar, C. Benjamin, \emph{The emergence of cooperation in the thermodynamic limit}, \href{https://doi.org/10.1016/j.chaos.2020.109762}{Chaos \textbf{135}, 109762 (2020).}

\bibitem{ref6}S. Galam, B. Walliser, \emph{Ising model versus normal form game}, \href{https://doi.org/10.1016/j.physa.2009.09.029}{Physica A: Statistical Mechanics and its Applications, \textbf{389:3} (2010).}

\bibitem{ref6a}M. Nowak, K. Sigmund, \emph{Tit for tat in heterogeneous populations} \href{https://doi.org/10.1038/355250a0}{Nature \textbf{355}, 250–253 (1992)}. 

\bibitem{ref6b}M. Nowak, R. May, \emph{Evolutionary games and spatial chaos},\href{https://doi.org/10.1038/359826a0}{Nature \textbf{359}, 826–829 (1992)}. 

\bibitem{ref6c}M. Nowak, K. Sigmund, \emph{A strategy of win-stay, lose-shift that outperforms tit-for-tat in the Prisoner's Dilemma game}, \href{https://doi.org/10.1038/364056a0}{Nature 364, 56–58 (1993)}.


\bibitem{ref7}S. Sarkar, C. Benjamin, \emph{Quantum Nash equilibrium in the thermodynamic limit}, \href{https://doi.org/10.1007/s11128-019-2237-2}{Quantum Inf Process \textbf{18}, 122 (2019).}

\bibitem{ref8}J. Eisert, M. Wilkens, M. Lewenstein, \emph{Quantum Games and Quantum Strategies}, \href{https://doi.org/10.1103/PhysRevLett.83.3077}{Phys. Rev. Lett. \textbf{83}, 3077 (1999).}


\bibitem{ref9}C. Benjamin, S. Sarkar, \emph{Triggers for cooperative behaviour in the thermodynamic limit: A case study in Public goods game}, \href{https://doi.org/10.1063/1.5085076}{Chaos \textbf{29}, 053131 (2019).}

\bibitem{ref10}C. Benjamin, Arjun Krishnan U.M., \emph{Nash equilibrium mapping vs. Hamiltonian dynamics vs. Darwinian evolution for some social dilemma games in the thermodynamic limit}, \href{https://doi.org/10.1140/epjb/s10051-023-00573-4}{Eur. Phys. J. B \textbf{96}, 105 (2023).}

\bibitem{ref11}C. Adami, A. Hintze, \emph{Thermodynamics of evolutionary games}, \href{https://journals.aps.org/pre/abstract/10.1103/PhysRevE.97.062136}{Phys. Rev. E \textbf{97}, 062136 (2018).}

\bibitem{ref12}R. Tah, C. Benjamin, \emph{Game susceptibility, Correlation and Payoff capacity as a measure of Cooperative behaviour in the thermodynamic limit of some Social dilemmas}, \href{https://doi.org/10.48550/arXiv.2401.18065}{arXiv:2401.18065 [cond-mat.stat-mech] (2024).}

\bibitem{ref12a}R. Tah, C. Benjamin, \emph{Agent-based Modelling of Quantum Prisoner's Dilemma}, \href{https://doi.org/10.48550/arXiv.2404.02216}{arXiv:2401.18065 [arXiv:2404.02216 (2024).}

\bibitem{ref13}P.M. Altrock, A. Traulsen, \emph{Deterministic evolutionary game dynamics in finite populations}, \href{https://journals.aps.org/pre/abstract/10.1103/PhysRevE.80.011909}{Phys. Rev. E \textbf{80}, 011909 (2009).}

\bibitem{ref14} G.T. Landi, \emph{Ising model and Landau theory, Undergraduate Statistical Mechanics, IFUSP - Physics Institute, University of São Paulo}, \href{http://www.fmt.if.usp.br/~gtlandi/lecture-notes/12---ising.pdf}{Chapter \textbf{3}, Ferromagnetism (2017).}

\bibitem{ref15} C. Benjamin, A. Dash, \emph{Switching global correlations on and off in a many-body quantum state by tuning local entanglement}, \href{https://doi.org/10.1063/5.0171825}{Chaos \textbf{33}, 091104 (2023).}

\bibitem{ref16} M. DeVos, D.A. Kent, \emph{Game Theory: A Playful Introduction}, \href{https://bookstore.ams.org/stml-80}{American Mathematical Society (2016).}

\bibitem{ref17} A.P. Flitney, D. Abbott, \emph{An Introduction to Quantum Game Theory}, \href{https://doi.org/10.1142/S0219477502000981}{Fluctuation and Noise Letters, \textbf{02}, \textbf{R175-R187} (2002).}

\bibitem{ref18} S. Schecter, H. Gintis, \emph{Game Theory in Action: An Introduction to Classical and Evolutionary Models}, \href{https://press.princeton.edu/books/hardcover/9780691167640/game-theory-in-action}{Princeton University Press (2016).}

\bibitem{ref19} S. Sarkar, C. Benjamin, \emph{Entanglement renders free riding redundant in the thermodynamic limit}, \href{https://doi.org/10.1016/j.physa.2019.01.085}{Physica A: Stat. Mech. and App., \textbf{521}, 607-613 (2019).}

\bibitem{ref20} G. Jaeger, \emph{The Ehrenfest Classification of Phase Transitions: Introduction and Evolution}, \href{https://doi.org/10.1007/s004070050021}{Arch Hist Exact Sc. \textbf{53}, 51–81 (1998).}

\bibitem{ref21} A. Sandvik, \emph{Computational Physics lecture notes, Fall 2023, Department of Physics, Boston University}, \href{https://physics.bu.edu/~py502/slides/l17.pdf}{Measuring physical observables.}

\bibitem{ref22} G. Hocky \textit{et al.}, \emph{Statistical Mechanics lecture notes, 2021, New York University}, \href{https://hockygroup.hosting.nyu.edu/exercise/ising-1d.html}{Metropolis Monte Carlo for the Ising Model.}

\bibitem{ref23} R.J. Baxter, \emph{Exactly Solved Models in Statistical Mechanics, 1982, Academic Press.}

\bibitem{ref24} A. Bordg, Y. He, \emph{Comment on ``Quantum Games and Quantum Strategies"}, \href{https://doi.org/10.48550/arXiv.1911.09354}{arXiv:1911.09354 [quant-ph] (2019)}

\bibitem{ref25} De He, T. Ye, \emph{An Improvement of Quantum Prisoners’ Dilemma Protocol of Eisert-Wilkens-Lewenstein}, \href{https://doi.org/10.1007/s10773-019-04351-w}{Int J Theor Phys \textbf{59}, 1382–1395 (2020).}

\bibitem{ref26} \href{https://github.com/rajdeep2810/RCPD---Agent-based-modelling}{GitHub link for Agent-based modelling codes for RPD.}


\end{thebibliography}
\end{document}